\documentclass[]{aa}  

\usepackage{graphicx}
\usepackage{txfonts}
\usepackage[breaklinks=true]{hyperref} 
\usepackage{natbib,twoopt}
\usepackage{float}
\usepackage{arydshln}
\usepackage[breaklinks=true]{hyperref} 
\bibpunct{(}{)}{;}{a}{}{,}                            
\usepackage{lscape}

\newcommand{\ra}{$\,\alpha$\xspace}
\newcommand{\dec}{$\,\delta$\xspace}
\newcommand{\pmra}{$\,\mu_{\alpha^{*}}$\xspace}
\newcommand{\pmdec}{$\,\mu_{\delta}$\xspace}
\newcommand{\para}{$\,\varpi$\xspace}
\newcommand{\Av}{$\,A_\mathrm{V}$\xspace}
\newcommand{\Ag}{$\,A_\mathrm{G}$\xspace}
\newcommand{\alphaIR}{$\,\alpha_\mathrm{IR}$\xspace}
\newcommand{\microm}{$\,$\mu\mathrm{m}$$\xspace}
\newcommand{\gaia}{{\it Gaia}\xspace}
\newcommand{\eps}{$\epsilon$\xspace}
\newcommand{\mpts}{{\it mPts}\xspace}
\newcommand{\mcls}{{\it mCls}\xspace}
\newcommand{\dbscan}{\texttt{DBSCAN}\xspace}
\newcommand{\optics}{\texttt{OPTICS}\xspace}
\newcommand{\hdbscan}{\texttt{HDBSCAN}\xspace}
\newcommand{\rhooph}{$\rho$ Oph\xspace}
\newcommand{\simbad}{\texttt{SIMBAD}\xspace}
\newcommand{\vizier}{\texttt{VizieR}\xspace}

\begin{document} 

\title{Census of $\rho$ Oph candidate members from \gaia Data Release 2\footnote{Tables 3, A.1, and A.4 are entirely available in electronic form at the CDS via anonymous ftp to cdsarc.u-strasbg.fr (130.79.128.5) or via http://cdsweb.u-strasbg.fr/cgi-bin/qcat?J/A+A/}
}


\author{H. C\'anovas\inst{1}, 
   C. Cantero\inst{1},
   L. Cieza\inst{2}, 
   A. Bombrun\inst{3}, 
   U. Lammers\inst{1}, 
   B. Mer\'in\inst{1}, 
   A. Mora\inst{4}, 
   \'A. Ribas\inst{5}, and
   D. Ru\'iz-Rodr\'iguez\inst{6}
}

\institute{ 
European Space Astronomy Centre (ESA/ESAC), Operations Department, Villanueva de la Ca\~nada (Madrid), Spain\\
\email{hcanovas@sciops.esa.int}
\and N\'ucleo de Astronom\'ia,  Facultad de Ingenier\'ia y Ciencias,  Universidad Diego Portales, Av. Ej\'ercito 441, Santiago, Chile
\and HE Space Operations B.V. for ESA, European Space Astronomy Centre (ESA/ESAC), Operations Department, Villanueva de la Ca\~nada (Madrid), Spain
\and Aurora Technology B.V. for ESA, European Space Astronomy Centre (ESA/ESAC), Operations Department, Villanueva de la Ca\~nada (Madrid), Spain
\and European Southern Observatory (ESO), Alonso de C\'ordova 3107, Vitacura, Casilla 19001, Santiago de Chile, Chile
\and Chester F. Carlson Center for Imaging Science, School of Physics $\&$ Astronomy, and Laboratory for Multiwavelength Astrophysics,
Rochester Institute of Technology, 54 Lomb Memorial Drive, Rochester NY 14623 USA
}

\date{\today}

\abstract
{The Ophiuchus cloud complex is one of the best laboratories to study the earlier stages of the stellar and protoplanetary
disc evolution. The wealth of accurate astrometric measurements contained in the \gaia Data Release 2 can be
used to update the census of Ophiuchus member candidates.}
{We seek to find potential new members of Ophiuchus and identify those surrounded by a circumstellar disc.}
{We constructed a control sample composed of 188 bona fide Ophiuchus members. Using this sample as a reference
we applied three different density-based machine learning clustering algorithms (\dbscan, \optics, and \hdbscan)
to a sample drawn from the \gaia catalogue centred on the Ophiuchus cloud. The clustering analysis was applied in
the five astrometric dimensions defined by the three-dimensional Cartesian space and the proper motions in right
ascension and declination.}
{The three clustering algorithms systematically identify a similar set of candidate members in a main cluster with
astrometric properties consistent with those of the control sample. The increased flexibility of the \optics and \hdbscan
algorithms enable these methods to identify a secondary cluster. We constructed a common sample containing
391 member candidates including 166 new objects, which have not yet been discussed in the literature. By combining the \gaia data
with 2MASS and WISE photometry, we built the spectral energy distributions from $0.5 \microm$ to $22\microm$ for a subset
of 48 objects and found a total of 41 discs, including 11 Class II and 1 Class III new discs. }
{Density-based clustering algorithms are a promising tool to identify candidate members of star forming regions in
large astrometric databases. By combining the \gaia data with infrared catalogues, it is possible to discover new
protoplanetary discs. If confirmed, the candidate members discussed in this work would represent an increment of roughly
$40-50\%$ of the current census of Ophiuchus.}


\titlerunning{Census of $\rho$-Oph candidate members from \gaia DR2}
\authorrunning{Canovas et al.}
\maketitle

\section{Introduction}

Star forming regions (SFR) composed of hundreds of pre-main sequence (PMS) stars are natural laboratories to
learn about the early stages of the stellar evolution process. These places are crucial to study the birth sites of planets
as a significant fraction of PMS stars are surrounded by protoplanetary discs. Understanding and identifying the
mechanisms driving the evolution of these discs is key to explain how planetary systems are formed \citep[e.g.][]
{2016JGRE..121.1962M}. With this goal in mind several teams have observed large populations of protoplanetary
discs in various SFRs across a range of wavelengths \citep[e.g. ][]{Cieza_2018, Dent_2013,
2009ApJS..181..321E, 2001ApJ...553L.153H}. Analyses of extensive disc samples (and their stellar hosts)
have revealed trends such as the disc mass-stellar mass relation \citep{2013ApJ...771..129A, 2016ApJ...828...46A,
2016ApJ...831..125P} and the decrement of disc mass in millimeter-sized grains with stellar age \citep{2017AJ....153..240A,
Ru_z_Rodr_guez_2018}. A general limitation of these studies is the lack of a complete census of the members of
the region under analysis. Such a census is also desirable (among many other reasons) to assess the impact of the
environment over the observed disc properties because, for instance, stellar fly-bys are expected to affect disc sizes,
shapes, and masses \citep{Bate_2011, Bate_2018, 2019MNRAS.483.4114C}.

Identifying the members of any SFR is a difficult task hindered by various causes. Nearby (<400 pc from Earth)
regions generally occupy large areas ($\geq1$deg$^2$) on the projected sky, and therefore it is observationally
expensive to study these regions with classical observatories. Furthermore, the late spectral type objects that populate these
regions are difficult to detect and characterise as a consequence of their intrinsic lower luminosity, and the same applies to any
young stellar object (YSO) that has a very high extinction. Early works took advantage of the strong X-ray activity
and accretion driven $\mathrm{H}_{\alpha}$ emission of PMS stars to detect these objects using X-ray observatories and
$\mathrm{H}_\alpha$ objective prism surveys \citep[e.g.][]{1983ApJ...269..182M, Walter_1994, 1987AJ.....94..106W}.
More recently, the use of astrometric measurements has made it possible to identify or confirm hundreds
of members of several SFRs and young associations using different methods \citep{Bruijne_1999,
2013ApJ...762...88M, 1999AJ....117..354D}. The advent of the \gaia mission \citep{2016A&amp;A...595A...1G} is
revolutionising this (and many other) field(s), allowing for the detection of new candidate members of different stellar
populations by means of accurate astrometric measurements. In a few remarkable cases, a direct inspection by
eye of the \gaia astrometry can even reveal the members of a stellar cluster \citep[e.g. the Pleiades example in][]
{gaia_dr1_2016,2017arXiv170702160T}. However, in practice complex algorithms are needed to identify the
member candidates based on the astrometric properties of the studied sample \citep[e.g. ][]{2018ApJ...856...23G}.
This becomes increasingly difficult when processing large catalogues such as the \gaia Data Release 2
\citep[DR2;][]{Lindegren_2018}.

The so-called unsupervised machine learning (ML) clustering algorithms are a collection of software tools developed
to identify patterns or clusters in unlabelled databases. Within these algorithms there are multiple approaches to
the problem of cluster detection, such as centroid-based algorithms \citep[e.g. the $k$-means algorithm; ][]{kmeans},
distribution-based clustering (e.g. Gaussian-mixture models), or density-based algorithms \citep[for an overview
of clustering analysis in astronomy see e.g.][their Chapter 3.3 and references therein]{2012msma.book.....F}.
The latter are well-suited to identify arbitrarily shaped clusters that can be broadly described as overdensities in
a lower density space. An advantage of the density-based algorithms is that no prior knowledge about the analysed
dataset is needed. In other words, the user does not need to know the number of clusters present in the dataset,
and these algorithms do not assume any particular distribution (such as one or multiple Gaussians) when associating
the data points with a cluster. Density-based spatial clustering of applications with noise \citep[\dbscan;][]
{Ester_1996} stands out as one of the most famous algorithms in many disciplines, and it is becoming popular
in astronomy \citep[e.g.][]{Beccari_2018, Bianchini_2018, 2008AN....329..801C, Castro_Ginard_2018, Joncour_2018,
2019MNRAS.484..574H, Tramacere_2013, 2018A&A...618A..12W}. There are a number of modifications or
improvements of \dbscan and, among those, the ordering points to identify the clustering structure \citep[\optics;][]
{Ankerst_1999} and the hierarchical density-based spatial clustering of applications with noise \citep[\hdbscan;][]{hdbscan_1}
algorithms are becoming popular owing to their recognised performance to detect various types of clusters. However,
to date their use in the astronomical literature is still negligible except for a few cases \citep[e.g. ][]{2017MNRAS.465.3879C,
2017MNRAS.466.1648K, 2018MNRAS.475.4617K, 2017A&A...599A.143S}. 

In this paper we use the \dbscan, \optics, and \hdbscan algorithms to identify potential members of the Ophiuchus
SFR (hereafter \rhooph) in a sample drawn from the \gaia DR2 catalogue \citep{2018}. With average
distance and age estimates ranging from 120 pc to 140 pc and 2 Myr to 5 Myr, respectively \citep[][and references therein]
{2008hsf2.book..351W}, \rhooph has been the subject of numerous studies \citep[e.g. ][]{2009ApJ...700.1502A,
2007ApJ...671.1800A, Cheetham_2015, 2017ApJ...851...83C, 2011AJ....142..140E, Mamajek_2008, 2008hsf2.book..351W}.
Recently, 289 discs in this cloud have been observed with ALMA aiming to study in detail the different evolutionary
stages of the protoplanetary discs \citep{Cieza_2018}. 
The \rhooph region is located at the foreground of the southeastern edge of the Upper Scorpius (hereafter USco)
subgroup of the large Sco-Cen OB association \citep{1989A&A...216...44D, 1999AJ....117..354D}. Both USco and
\rhooph have similar proper motion distributions \citep{Mamajek_2008}, and it has been proposed that the star
formation in \rhooph was triggered by a supernova event in USco \citep{1992A&A...262..258D, 2011AJ....142..140E,
2002AJ....124..404P}. A major difference between both regions is their age and evolutionary stage. While the \rhooph
region is rich in molecular gas, has a very high optical extinction, and star formation is still ongoing, the opposite is
found in the $\sim 10$ Myr old USco \citep{2012ApJ...746..154P, 2016MNRAS.461..794P, 2008hsf2.book..351W}.
Given the youth of \rhooph, its members have similar velocity distributions and they are concentrated
in a relatively compact region of the Galaxy. In other words, the cloud members should appear clustered
in the multi-dimensional space defined by their spatial coordinates and kinematic parameters when compared
to the field stellar population. We applied the clustering algorithms in the five-dimensional space defined by
the three spatial coordinates and two kinematical parameters given by the proper motions on right ascension
\pmra\footnote{
\pmra = $\mu_{\alpha} \cos \delta$, where $\mu_{\alpha} = \frac{\alpha_1 - \alpha_2}{\Delta t}$ is the apparent
movement in right ascension $\alpha$ in a given time interval $\Delta t$ and $\delta$ is the object declination.}
and declination \pmdec. By comparing their results we aim to reduce the selection biases inherent to each
algorithm and therefore construct a robust sample of \rhooph members candidates. The samples used
for our study are described in Section 2. In Section 3 we describe our methodology and apply the three
algorithms to our \gaia sample. This section finishes by presenting a sample of sources simultaneously identified
by the three algorithms. In Section 4 we discuss the properties of this sample and we use infrared photometry
to identify a set of objects showing warm dust emission associated with a circumstellar disc. A summary and
conclusions are given in Section 5, while extra figures and tables are shown in the Appendix.

\section{Sample construction}
\label{sect_sample_selection}
\subsection{Initial sample}
\label{sect_sample_initial}
We begun by compiling a list of \rhooph member candidates identified as such by means of optical spectroscopy,
X-ray emission, or Spitzer photometry. To do so we considered the 316 objects listed by \cite{2008hsf2.book..351W},
the sample discussed by \cite{2011AJ....142..140E} (135 objects), and the catalogue of \rhooph YSO candidates 
observed by {\it Spitzer} during the Cores to Disks Legacy program \citep{2003PASP..115..965E} and presented
by \cite{Dunham_2015} (292 objects). The on-line versions of these catalogues, hosted by the \vizier service,
include a \simbad \citep{2000A&AS..143....9W} ID associated with each object. We used these IDs to obtain
the corresponding Two Micron All Sky Survey \cite[2MASS; ][]{2006AJ....131.1163S} IDs. There are no 2MASS
counterparts for 2 objects in the \cite{2008hsf2.book..351W} catalogue, 1 object from \cite{2011AJ....142..140E},
and 10 objects from \cite{Dunham_2015}, and these 13 sources are not considered in our study. After accounting
for duplicates we obtained our  initial sample, which contains 465 objects. This sample is listed in Table~\ref{tab:ap_tb_1}
in the Appendix.

\subsection{Control sample}
\label{sect_sample_control}
With the 2MASS IDs of our initial sample at hand we run an identity cross-match Astronomical Data Query
Language \citep[ADQL;][]{2008ivoa.spec.1030O} query between our initial sample and the \gaia DR2 catalogue
using the 2MASS-\gaia matched table (\texttt{gaiadr2.tmass\_best\_neighbour}) available on the \gaia archive\footnote{
https://www.cosmos.esa.int/web/gaia/data-release-2}.
This approach is advantageous compared to a sky cross-match by coordinates as there is no need to transform the
2MASS coordinates from the J2000 to the J2015.5 epoch\footnote{The astrometric source parameters in the \gaia
DR2 are referred to the J2015.5 epoch. Throughout this paper we use the same convention. For a detailed description
on coordinates transformations, see for example Sect. 3.1.7 in the \gaia DR2 on-line documentation at
https://gea.esac.esa.int/archive/documentation/GDR2/}.
Our query returned 304 sources. The missing objects are generally fainter at $J$ band (and therefore probably
fainter at $G$ band) than the detected objects (see Fig.~\ref{fig:ap_1} in the Appendix). The results listed on the
DR2 catalogue (such as the right ascension \ra, the declination \dec, the parallax \para, and the proper motions
in right ascension and declination \pmra and \pmdec) are obtained through a complex data analysis process that
relies on treating each observed source as a single, ``well behaved'' star \citep[the five-parameter model; see][]
{Lindegren_2012, Lindegren_2018}. Binaries and YSOs, which are typically variable and can appear as extended
sources at optical wavelengths, can be problematic for the astrometric solution. Therefore, we applied a
selection criteria to extract the sources with high quality astrometry. First, we removed the objects with parallax
signal to noise $\varpi/\sigma_{\varpi} < 10$ and visibility periods\footnote{
As explained in the DR2 archive documentation, a visibility period is a group of observations separated from other
groups by four or more days. See http://gea.esac.esa.int/archive/documentation/GDR2/index.html}
$\le7$ to produce an astrometrically precise and reliable dataset \citep[see][]{Lindegren_2018, Arenou_2018}.
Second, we only extracted the sources whose observations are consistent with the five-parameter model. To do so,
we applied the current quality criteria described in the \gaia technical note GAIA-C3-TN-LU-LL-124-01\footnote{
see https://www.cosmos.esa.int/web/gaia/dr2-known-issues and http://www.rssd.esa.int/doc\_fetch.php?id=3757412}
that is based on an empirical analysis of the \gaia DR2 data. This analysis introduces the renormalised unit
weight error (RUWE), which is a reliable indicator of the goodness-of-fit of the astrometric solution. Using the
RUWE as a quality index is especially useful for samples containing very red stars, such as the low-mass and extinct
PMS stars in \rhooph. We extracted the sources with RUWE $<1.40$ as recommended by this study.

This astrometrically cleaned sample contains 197 objects. We generated its histogram distributions over \para,
\pmra, and \pmdec, applying the Bayesian approach derived by \cite{2006physics...5197K} that optimises the
bin widths based on the data distribution and that is implemented in \texttt{Astropy} \citep{2013}. Throughout this
paper we use mas and mas yr$^{-1}$ as the reference units for the parallaxes and proper motions, respectively,
and all the histograms are constructed by applying the same methodology. These histograms show bell-shaped
distributions with a few objects located far away from the histogram peaks (Fig.~\ref{fig:ap_2}). We removed
these outliers by considering only the objects with $5<$\para$<9$ and $-20<$\pmra$<0$. The 9 objects excluded
by this filter are listed in Table~\ref{tab:ap_tb_2} in the Appendix. Given their parallaxes and proper motions
we consider these as background objects instead of \rhooph members. Our final sample, hereafter called control
sample, contains 188 targets. There are \gaia radial velocities for 7 of the objects, which have an average
value of $v_\mathrm{rad} = -7.7 \pm 2.4\,\mathrm{km\,s}^{-1}$. This sample occupies a sky region extending
from $[245.3\degr:249.9\degr]$ in right ascension and $[-25.4\degr:-22.9\degr]$ in declination, and has an
average parallax of \para$=7.1\pm0.4$. The average astrometric properties of the control sample are listed in
Table~\ref{tab:tb_1} and its parallax and proper motion histograms are shown in the Appendix (Fig.~\ref{fig:ap_3}).
Following \cite{2015PASP..127..994B} we computed the individual distances as $d = 1/\varpi$ since the parallax
fractional error of this sample is lower than 10. This way we obtained an average distance of $d = 141.2_{-7.5}^{+8.4}$
pc to this sample, where the uncertainty is dominated by the intrinsic dispersion of the dataset. In short, this
sample contains 188 bonafide members of the \rhooph cloud that can be used as a reference sample. The
control sample members are labelled in Table.~\ref{tab:ap_tb_1} with a  "Y" in the control column.
\begin{table}[ht!]
\begin{center}
\begin{tabular}{cccccc}
\hline
\hline
Stats & \ra & \dec & \para & \pmra & \pmdec\\
 & [$^{\circ}$] & [$^{\circ}$] & [$\mathrm{mas}$] & [$\mathrm{mas\,yr^{-1}}$] & [$\mathrm{mas\,yr^{-1}}$] \\
\hline
Mean & 246.8 & -24.3 & 7.1 & -7.2 & -25.5 \\
Sigma & 0.6 & 0.5 & 0.4 & 2.0 & 1.7 \\
\hline
\end{tabular}
\caption{Mean and standard deviation ($1\sigma$) of the control sample.}
\label{tab:tb_1}
\end{center}
\end{table}

\subsection{{\it Gaia} sample}
\label{sect_sample_gaia}

\begin{figure}[ht!]
  \centering \includegraphics[width=\columnwidth, trim = 0 40 0 0]{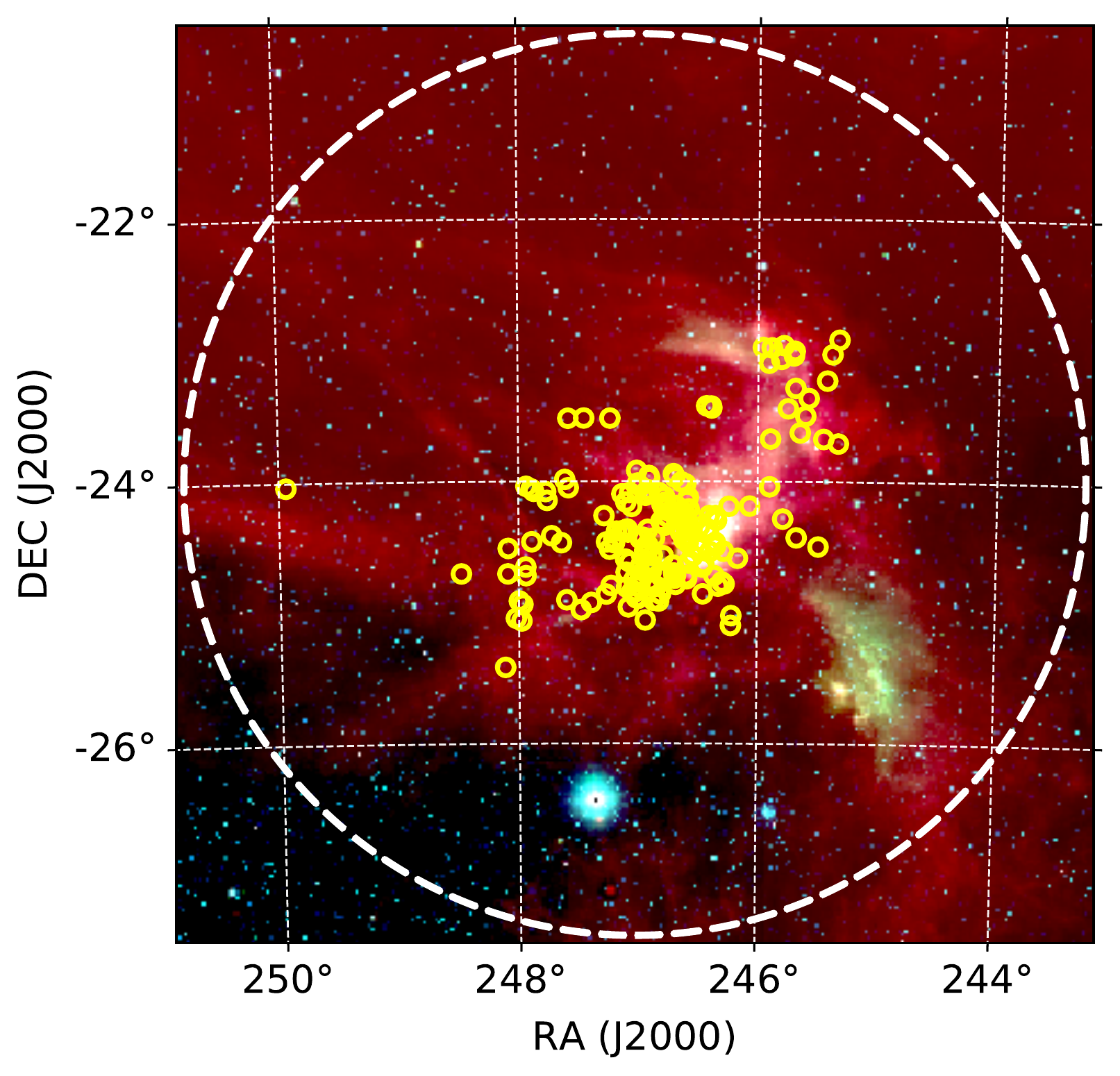}
  \caption{WISE image of the \rhooph region in a square-root stretched scale with RGB colours mapped to 22,
  4.6, and 3.4 $\mu$m. The control sample members are represented as yellow rings. The white dashed ring encompasses
  the area queried in the \gaia server. The saturated blue star towards the south is $\alpha$ Scorpii (a.k.a. Antares).}
   \label{fig:fig_1}
\end{figure}

\begin{figure}[ht!]
  \centering \includegraphics[width=1.0\columnwidth, trim = 0 30 0 0]{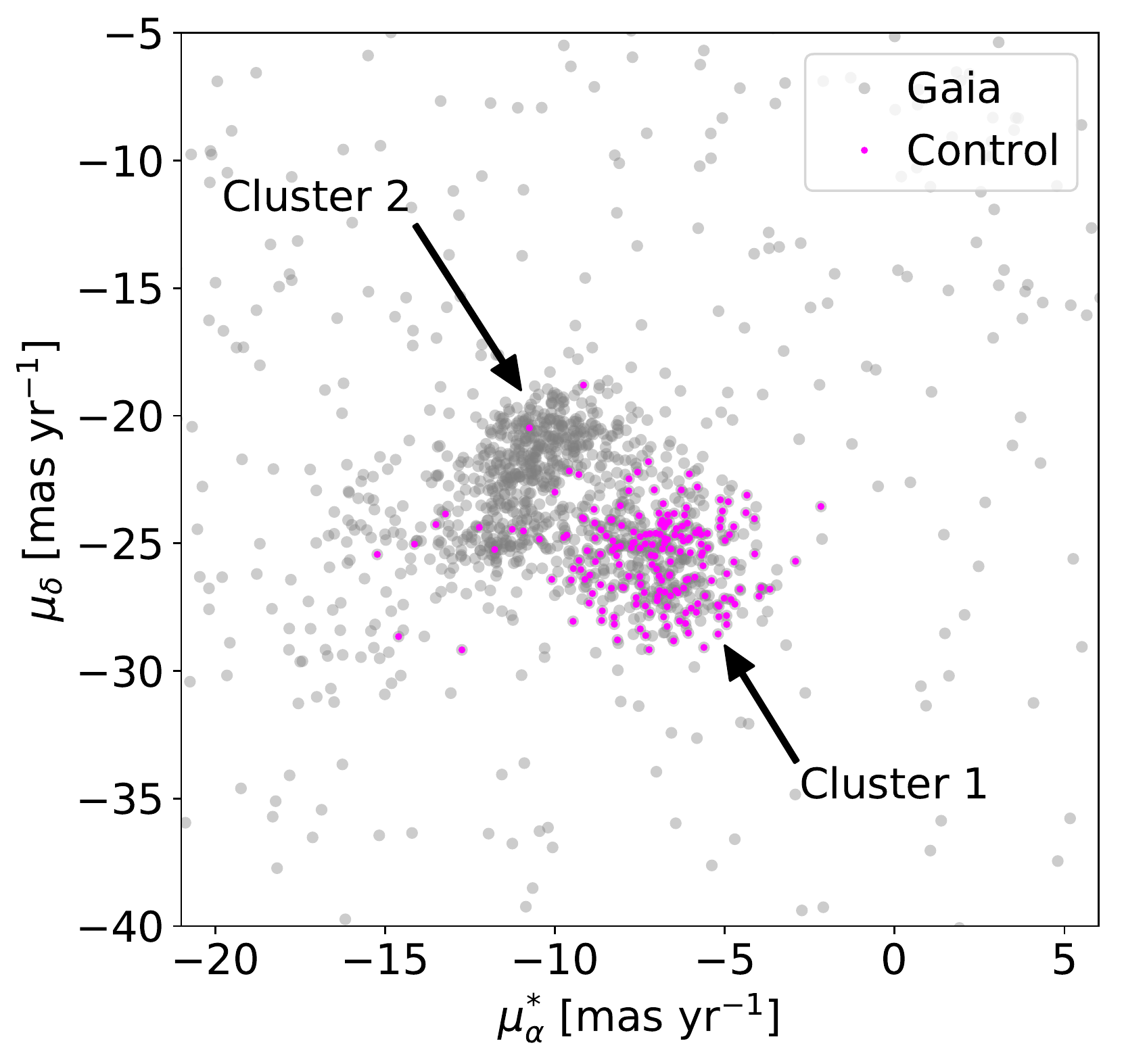}
  \caption{Zoom-in of the proper motions of the \gaia sample represented with grey circles. Their diameter roughly 
  equals the average error on \pmra (0.35 mas yr$^{-1}$), which is larger than the average error in \pmdec (0.23 mas yr$^{-1}$).
  The control sample members are represented as small magenta circles.}
  \label{fig:fig_2}
\end{figure}

In order to create a large sample to search for potential new members of \rhooph we run an ADQL cone search on
the {\it Gaia} server centred at $[247.0\degr, -24.0\degr]$ in $[\alpha, \delta]$ with a search radius of $r = 3.5\degr$
chosen to generously encompass the control sample (see Fig.~\ref{fig:fig_1}). We imposed a parallax range from
$[5:9]$ and parallax signal to noise S/N $>10$. This query returned 2814 targets that were reduced to 2300, including
the 188 objects of the control sample, after applying the quality selection criteria described in Sect.~\ref{sect_sample_control}.
This sample, labelled hereafter as the \gaia sample, contains radial velocity values for 236 sources with an average
value of $v_\mathrm{rad} = -6.9\pm31.8\,\mathrm{km\,s}^{-1}$. Figure~\ref{fig:fig_2} shows a zoom in of the \pmra versus
\pmdec of the \gaia sample; the control sample members are overlaid on top as small magenta circles. There is an evident
overdensity of sources around the centre of the plot. This overdensity seems to contain two different regions or clusters:
one with most of its members located around  $[-7.5, -25]$, and the second roughly centred at  $[-12, -21]$ in
[\pmra, \pmdec]; these are labelled as ``Cluster 1" and ``Cluster 2" in Fig.~\ref{fig:fig_2}, respectively. The control sample
members are mostly concentrated in Cluster 1.


\section{Analysis}
\label{sect_analysis}

\subsection{Basic concepts of density-based clustering}
\label{sect_methodology_2}
It is not straightforward to discriminate between the diffuse cluster(s) with no sharp boundaries and the population of
background or foreground objects shown in Fig.~\ref{fig:fig_2}. As mentioned in the introduction, there are multiple software tools
that are able to identify clusters embedded in large databases. We chose to use the family of density-based clustering algorithms
because they are especially well suited to detect arbitrarily shaped clusters, and by construction these algorithms
are not limited to detect clusters having a particular data distribution such as a Gaussian. We applied and compared
three different density-based clustering algorithms: \dbscan, \optics, and \hdbscan. To date \dbscan is one of the most
popular density-based clustering algorithms and its strong impact on the data mining research community is well
recognised\footnote{https://www.kdd.org/News/view/2014-sigkdd-test-of-time-award\#}. 
Since its original publication by \cite{Ester_1996} several clustering algorithms have been developed aiming to improve
its performance and currently there are many options available in the related literature. In this work we chose to use the \optics
and \hdbscan algorithms because their clustering analysis is based on the hierarchical density structure of the data,
which improves the performance of \dbscan when analysing datasets with strong density gradients. The three algorithms
that we used can be freely downloaded from the web repositories that we provide in the next subsections.

For our study we considered three spatial and two kinematic dimensions, with the former defined in Cartesian coordinates as 
\begin{eqnarray}
X &= & d\cdot \cos{\delta} \cos{\alpha}
\label{eq:eq_X} ,\\
Y &= & d\cdot \cos{\delta} \sin{\alpha}
\label{eq:eq_Y} ,\\
Z &= & d\cdot \sin{\delta}
\label{eq:eq_Z}
,\end{eqnarray}
where $d$ is the distance computed as the inverse of the parallax. Given the low fraction of objects with radial velocity
measurements in our \gaia sample we restricted the kinematic dimensions to the proper motions \pmra and \pmdec. Before
applying the clustering analysis the input dataset must be normalised to ensure that the data dispersion across the
processed dimensions is comparable. To do so we used the {\it Standard Scaler} (with an Euclidean metric)
implemented in the Python Machine Learning library scikit-learn \citep{scikit-learn}, ensuring that the mean and
variance of the data across each dimension are 0 and 1, respectively.

\begin{figure}[ht!]
  \centering \includegraphics[width=0.9\columnwidth, trim = 0 0 0 0]{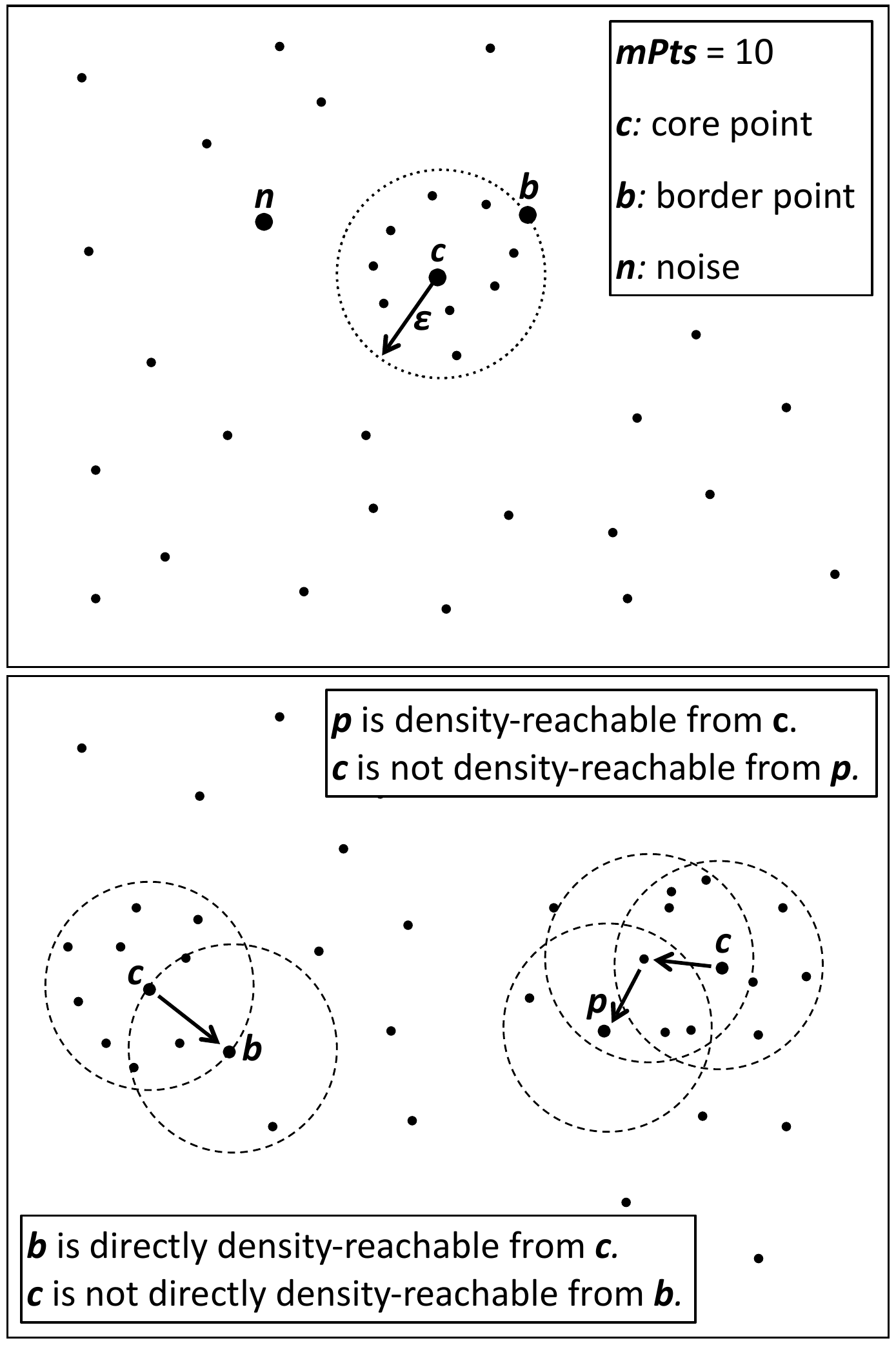}
  \caption{Fundamental concepts used by \dbscan and \optics.}
   \label{fig:fig_3}
\end{figure}

In the following lines we introduce the fundamental concepts used by the \dbscan, \optics and \hdbscan algorithms.
Clusters can be broadly described as localised and arbitrarily shaped regions of an $N$-dimensional space with an
excess of points per volume unit. In other words, the density in the neighbourhood of each cluster point must exceed
some threshold value. The points that do not satisfy this condition do not belong to the cluster and are classified as
noise. Two hyperparameters\footnote{
In ML jargon these are the parameters specified by the user before the clustering algorithm begins the learning process.}
can be used to describe this density threshold: first, \eps or EPS-distance, which is the distance between two points
in the $N$-dimensional space, and second, \mpts, which is the minimum amount of points that are needed to form a cluster.
A cluster can contain core and border points and by definition a cluster must contain at least one core point. The cores
are the points having at least \mpts neighbours within a distance $d = $ \eps. Border points are separated by a
distance $d \leq$ \eps from a core point; they are  directly density-reachable. Border points also have a number of neighbours
$n<$ \mpts\;; i.e. they are less dense than core points. Points separated by distances $d \gg$ \eps belong to the
same cluster if they are density-reachable, that is, if there is a chain of points $p_1, ..., p_n$ such that $p_{i+1}$
is directly density-reachable from $p_i$. Figure~\ref{fig:fig_3} illustrates these concepts. Below we present the results
of our analysis starting with a basic description of each algorithm. For a thorough description of the clustering algorithms
we refer to the specialised literature \citep[e.g.][]{Ankerst_1999, hdbscan_2, Ester_1996}.

\subsection{DBSCAN}
\label{sect3_dbscan}

The \dbscan\footnote{We use the scikit implementation by \cite{scikit-learn}.}
code is one of the most widely used density-based clustering algorithms. Originally introduced by \cite{Ester_1996}, this algorithm relies
on the \eps and \mpts hyperparameters to find arbitrarily shaped clusters of constant density. The algorithm results
strongly depend on these input parameters and the user may have to select the \mpts and \eps using a trial and error
approach.

We begin with the qualitative approach proposed by \cite{Ester_1996} to identify optimal \mpts and \eps values to find
clusters in the data. Their method consists in inspecting by eye the so-called sorted $k$-distance plot, which shows
the distance to the $k^{th}$ nearest neighbour (the $k$-distance) for each point with all the points sorted out by
decreasing $k$-distance. By construction, the points that belong to a cluster containing $k$ members have a lower
distance to their closest $k$-neighbours than the noise points. Therefore, a compact cluster containing $n \ge k$
points creates an abrupt decrement in its sorted $k$-distance plot. The $k$-distance at which the curve reduces
again its slope corresponds to the optimal \eps to identify a cluster using \mpts$= k$. This way, examining these curves
for different $k$ values allows us to identify pairs of \eps and \mpts hyperparameters. 

\begin{figure}[ht!]
  \centering \includegraphics[width=\columnwidth, trim = 0 100 0 0]{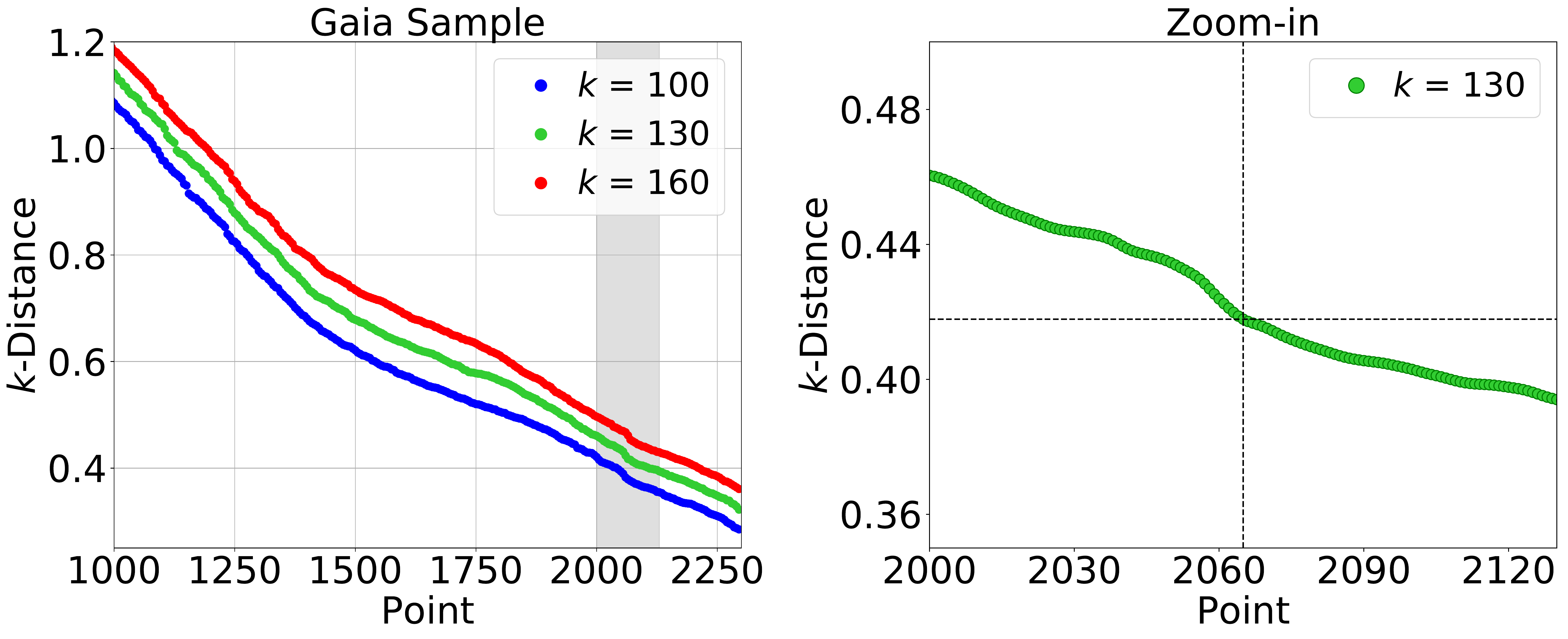}
  \caption{{\bf Left:} $k$-distance plots of the \gaia sample for different $k$ (or equivalently \mpts) values. The grey rectangle
  encompasses the step-like slope change in the curves (see text). {\bf Right:} Zoom-in of the $k = 130$ case with the bottom
  of the step-like slope change indicated by the dashed lines.}
  \label{fig:fig_4}
\end{figure}

We generated a set of $k$-distance curves for $k$ ranging between $[80:200]$. A careful look reveals an step-like
slope change located at point $\sim 2065$ for $k$ values ranging between $[100:160]$, as indicated by the grey
shaded region in the left panel of Fig.~\ref{fig:fig_4}. To find the appropriate \eps values for the different $k$ values
(and equivalently, for different \mpts) we computed the first and second order derivatives of the sorted $k$-distance
curves. Given the density distribution of our dataset we had to resample and smooth the
$k$-distance curves with a Gaussian kernel to isolate the $2^{nd}$-order derivative maxima with sampling and smoothing values varying
with \mpts. After finding the pairs of \eps and \mpts values for different \mpts we run \dbscan. The algorithm identifies
one single cluster dominated by a population of stars with astrometric distributions consistent with those of  our
control sample. The results of this exploration for three representative cases and their corresponding histogram
distributions in parallaxes and proper motions are listed in Table~\ref{tab:tb_2} and presented in Fig.~\ref{fig:fig_6}
(top row), respectively. The histograms in \pmra show a secondary peak at \pmra $\sim -12.5$, which becomes
increasingly significant with \mpts. A similar behaviour is observed in the \pmdec distribution, which shows a
secondary peak at \pmdec $\sim -23$ and is most evident for \mpts$=100$. The number of cluster elements
varies by $~\sim10\%$ within the explored range.

\subsection{OPTICS}
\label{sect3_optics}

By construction, all the clusters found by \dbscan in a given dataset have roughly the same density. Furthermore,
this algorithm struggles to identify all the  members in clusters with strong density gradients, such as a cluster composed of a very dense core surrounded by a low density ``halo''. The hierarchical clustering algorithm \optics\footnote{
We use the pyclustering (v0.8.1) implementation by \cite{andrei_novikov_2018_1254845}.}
\citep{Ankerst_1999} attempts to overcome these issues by focussing on the density-based clustering structure of
the data.\ The \optics algorithm constructs clusters of different densities by exploring a range of \eps values and working like an
expanded version of \dbscan.

As a first step, \optics finds the densest regions of the cluster and stores this information into two variables named
core distance and reachability distance. The former is the distance from a core point to its closest $mPts^{th}$
neighbour, and the latter is the smallest distance that makes a point  density-reachable from a core point. For
a given value of \mpts \optics classifies the points according to their reachability distance from the densest part of
the cluster. This is reflected in the so-called reachability plot, which shows a series of characteristic valleys, each
associated with a potential cluster. The bottom of the valley corresponds to the densest cluster region and the width of the valley
roughly scales with the number of cluster elements. The shape of the valley depends on the density distribution
of the cluster and the entire dataset. As a second step, \optics is executed with the \mpts and \eps hyperparameters
derived from the reachability curves as explained below.

Following \cite{Ankerst_1999} we explored the reachability curves for \mpts ranging between $[10:20]$. All the curves
show a wealth of substructure in the form of multiple local narrow valleys that smoothly disappear with increasing \mpts.
The curves also show a main and a secondary (less pronounced) valley (Fig.~\ref{fig:fig_5}, left panel). We did
a first exploration of the two clusters associated with these valleys (see below). The secondary valley is produced by a
cluster with $\sim180$ elements and has averages \para$=6.3\pm0.2$, \pmra$=-10.6\pm1.2$, and \pmdec$=-21.4\pm1.4$.
Its members are part of  Cluster 2 shown in Fig.~\ref{fig:fig_2} and only one of the control sample members is part
of this cluster. On average, the astrometric properties of this secondary cluster are different from those of the control
sample and, at first sight, they are consistent with a population of USco stars located in the background of our control
sample. We defer a detailed analysis of this cluster to a separate and dedicated study. In what follows we focus our
discussion on the main valley and its associated cluster, as this one has astrometric properties similar to those of the
control sample. In the explored range the shape of the valley is similar to the ``case B'' described by \cite{Ankerst_1999}
in their Sect. 4.3 (see also their Figs. 17 and 18). The valley shows a local plateau at its beginning (labelled as starting
of the cluster; SoC), which has similar y-coordinates as the cluster end; this end appears as an abrupt decrement in the
curve and is labelled as reachability end (Reach End in the right panel of Fig.~\ref{fig:fig_5}).
In this case, the appropriate \eps to extract the cluster for \mpts$=16$ corresponds to the $y-$coordinates of the
reachability end, that is, \eps$=0.259$. The beginning and end of the main valley are clearly detected for \mpts
ranging between $[14:18]$, and therefore we restrict our analysis to this range. This way we derived the \eps values
for the corresponding \mpts and then executed \optics using these hyperparameters. As representative cases we
consider those in which the reachability end is clearly detected. Those are listed in Table~\ref{tab:tb_2}, and their
corresponding histograms in\ parallax and proper motions are shown in Fig.~\ref{fig:fig_6} (central row). All the
cases produce a similar outcome with a variation in cluster elements $<10\%$. The histogram distributions show
a secondary peak at \pmra $\sim -12.5$ and \pmdec $\sim -23$.
\begin{figure}[ht!]
\centering \includegraphics[width=\columnwidth, trim = 0 100 0 0]{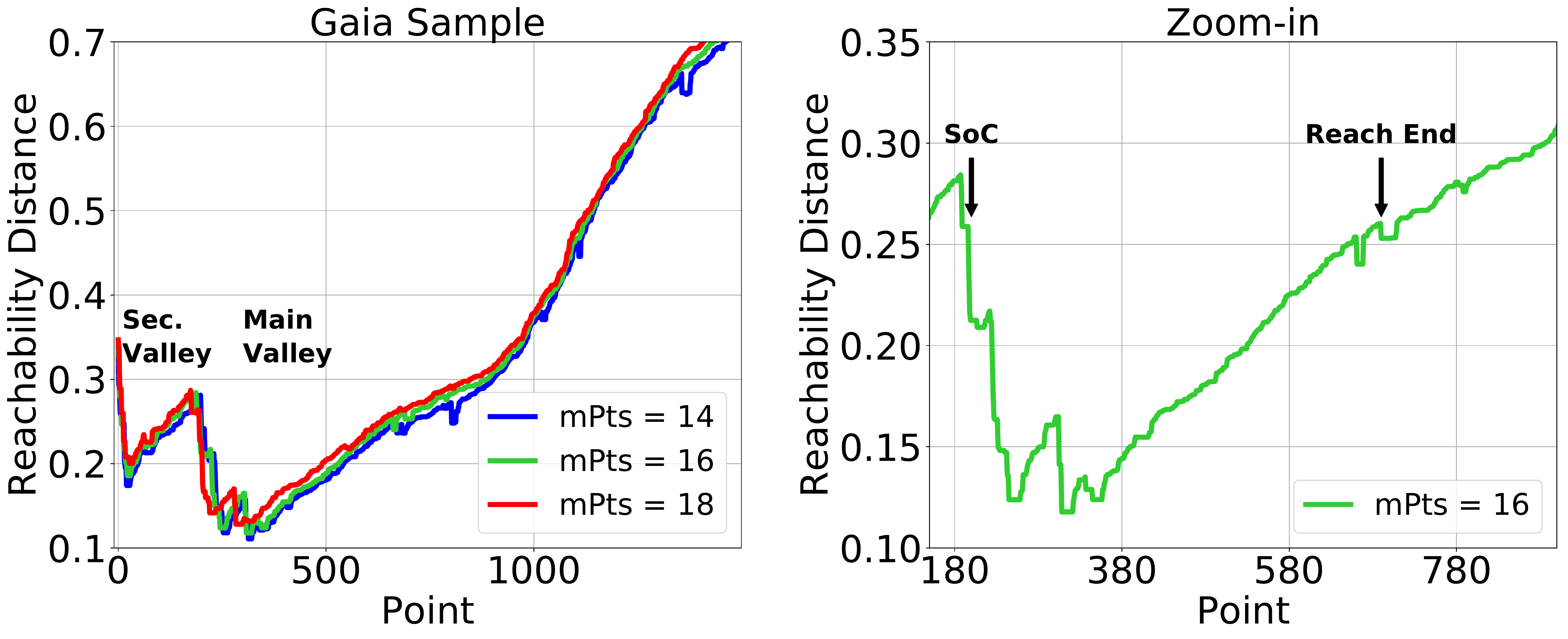}
\caption{{\bf{Left:}} Reachability distance plots of the \gaia sample computed for a range of \mpts. {\bf Right:} Zoom-in
of the \mpts$=16$ case, highlighting the location of the first and last points of the main valley.}
\label{fig:fig_5}
\end{figure}

\subsection{HDBSCAN}
\label{sect3_hdbscan}
A drawback of both \dbscan and \optics is the potential difficulty in finding optimal \eps and \mpts values. In high-density
datasets it can be complicated to identify unambiguously the step-like slope change in the $k$-distance curves used
by \dbscan and the first and last points of the valleys in the reachability-distance plots generated by \optics. The hierarchical
algorithm \hdbscan\footnote{We use the implementation by \citet{McInnes2017}.} 
\citep{hdbscan_1, hdbscan_2} requires one single hyperparameter to operate (the ``minimum cluster size'' \mcls,
conceptually similar to \mpts), and it therefore simplifies the task of finding appropriate hyperparameters. Similar to \optics,
\hdbscan can identify clusters with different densities and it is sensitive to the density gradients inside a cluster.

The \hdbscan algorithm is much more complex than \dbscan and \optics and therefore we briefly describe its main steps,
referring to \cite{hdbscan_2} for a thorough description of the algorithm. First, the low-density and noise points are identified
by computing the ``mutual reachability distance''. This quantity is the maximum of three distances for each pair of data
points $A$ and $B$: the core distance to the $k^{th}$ point for $A$ and $B$, and the metric distance between $A$ and
$B$. Second, \hdbscan uses the mutual reachability distance to identify and classify the densest points of the dataset
according to their relative density. The points are grouped into clusters through a multi-stage process that involves
generating the minimum spanning tree of the mutual reachability distances, computing the data points density hierarchy,
and creating a hierarchical condensed cluster tree. Apart from identifying the clusters in a given dataset, \hdbscan uses
the condensed cluster tree to assign a membership probability to each member of a cluster. 

\begin{figure}[ht!]
\centering \includegraphics[width=\columnwidth, trim = 0 100 0 0]{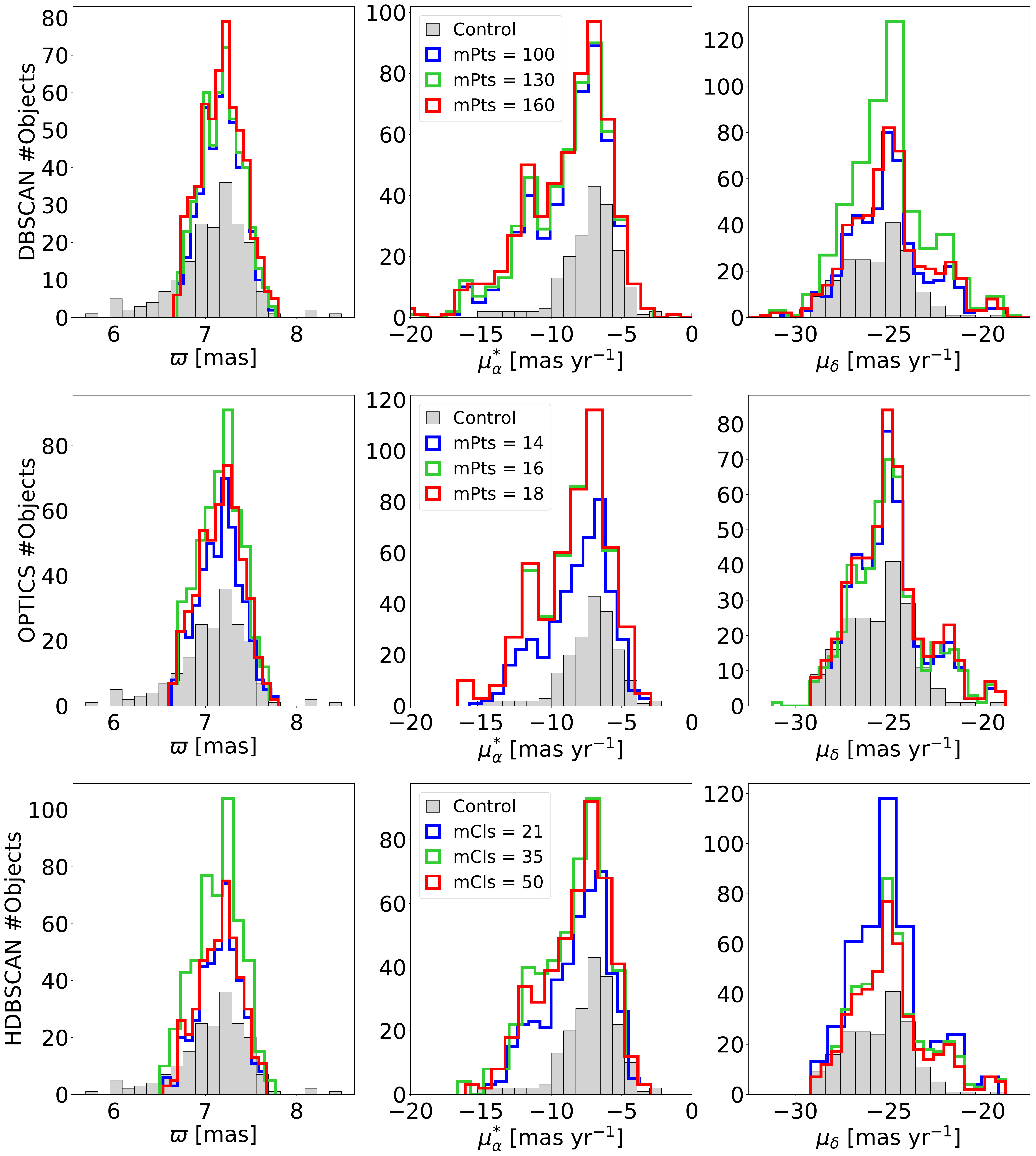}
\caption{Histogram distributions of the cluster identified by \dbscan (top) and the main cluster identified by
\optics (centre) and \hdbscan (bottom) for different hyperparameters (see legend and Table~\ref{tab:tb_2}).
The control sample is represented with grey bars.}
\label{fig:fig_6}
\end{figure}

We set the probability threshold to the maximum, i.e. we only considered cluster members with $100\%$ associated
membership probability, and we then explored the range of \mcls values from $[20:100]$. For \mcls $>58$ \hdbscan does
not find any cluster, while for \mcls between $[20:58]$ the algorithm systematically identifies a main and secondary
cluster. After a first exploration we find that the secondary cluster has between 30 and 50 elements, none of which are
included in the control sample. As with \optics, the members of this secondary cluster have proper motions consistent
with those of Cluster 2 in Fig.~\ref{fig:fig_2} (with averages \para$=6.3\pm0.1$, \pmra$=-10.7\pm0.9$, and \pmdec$=-21.3\pm1.3$). As
explained before, in this paper we focus our attention on the main cluster found by \hdbscan because its astrometric
properties are consistent with those of the control sample. For \mcls ranging from $[20:58]$ the number of cluster
members mostly oscillates around $~460$. As representative outputs we consider those listed in Table~\ref{tab:tb_2}
as, excluding a few cases with number of elements $<200$, those encompass the minimum, maximum, and roughly
average number of elements of the main cluster. The corresponding histograms are shown in Fig.~\ref{fig:fig_6}
(bottom row). The proper motion distributions show two secondary peaks at  $\sim -12.5$ in \pmra and $\sim -23$
in \pmdec, albeit these peaks are less pronounced than in the \dbscan and \optics cases.

\begin{table}[ht!]
\begin{center}
\begin{tabular}{ccccc}
\hline
Algorithm & mPts & Eps & Elements & Control (\%) \\
\hline
\dbscan & 100 & 0.373 & 492 & 155 (82.4) \\
\dbscan & 130 & 0.418 & 524 & 156 (83.0) \\
\dbscan & 160 & 0.455 & 552 & 158 (84.0) \\
\hline
\optics & 14 & 0.245 & 451 & 157 (83.5) \\
\optics & 16 & 0.259 & 494 & 157 (83.5) \\
\optics & 18 & 0.265 & 497 & 159 (84.6) \\
\hline
\hdbscan & 21 & -- & 427 & 153 (81.4) \\
\hdbscan & 35 & -- & 502 & 163 (86.7) \\
\hdbscan & 50 & -- & 462 & 158 (84.0) \\
\hline
\hline
\end{tabular}
\caption{Explored hyperparameters and number of cluster elements identified by each algorithm. The fifth column
indicates the number of control sample members found; the percentage is indicated inside a parenthesis. For
\hdbscan the first column corresponds to the \mcls.}
\label{tab:tb_2}
\end{center}
\end{table}

\subsection{Common sample}
\label{sect_common}

The three algorithms previously discussed find a population of objects with astrometric properties (excluding the yet
unknown radial velocity) consistent with those of the control sample, and additionally \optics and \hdbscan find a
secondary cluster with different average properties. As explained in Sect.~\ref{sect3_optics}, the fact that \dbscan does
not find this smaller cluster probably reflects one of the main weaknesses of this algorithm, which is its lack of sensitivity
to find clusters with different densities in the same dataset. Regarding the main cluster of our sample, the algorithm
outputs are slightly different and the cluster size is sensitive to the chosen algorithm and corresponding hyperparameters
(see Table~\ref{tab:tb_2}). On average, each algorithm recovers roughly 84\% of the control sample.
It is therefore not straightforward to decide which algorithm and hyperparameter combination produces the most reliable
sample of potential \rhooph members. To overcome this difficulty we merged the results listed in Table~\ref{tab:tb_2},
conservatively choosing the cluster combination that produces the smaller output in terms of elements. That is, we
combined the \dbscan output obtained for \mpts$= 100$, the \optics output with \mpts$= 14$, and the \hdbscan output
with \mcls$=21$. After accounting for duplicates the merged cluster (labelled hereafter as combined sample) contains
532 sources, 164 of which are included in the control sample. In this cluster there are 59 sources only detected by
\dbscan, and 13 sources only detected by \optics and other 13 only by \hdbscan. Table~\ref{tab:tb_3} lists the members
of this sample and includes a column labelled as "DOH" (acronym for \dbscan-\optics-\hdbscan), which indicates the
algorithm(s) that detect each member using a "Y/N" nomenclature. For example, if a source is detected only by \dbscan
its DOH value is "YNN", while the DOH\ value of a member detected by \optics and \hdbscan has is "NYY". From this
combined sample we extracted the sources simultaneously identified by the three algorithms (i.e. those with DOH$=$"YYY").
Hereafter we focus our analysis on this common sample that contains 391 sources, 148 of which belong to the control sample. 
\begin{table*}
\begin{center}
\begin{tabular}{cccccccc}
\hline
\ra & \dec & \para & \pmra & \pmdec & \texttt{source\_id} & Control & DOH \\
$[\mathrm{deg}]$ & $[\mathrm{deg}]$ & $[\mathrm{mas}]$ & [mas yr $^{-1}]$ & [mas yr$^{-1}]$ &  &  &  \\
\hline
\hline
243.3719939 & -23.1855471 & 7.2072955 & -8.8893582 & -25.4157242 & 6242554649124043520 & N & YNN \\
243.7959809 & -23.3786153 & 7.1327026 & -15.5389965 & -24.4785008 & 6050385511522948096 & N & YNN \\
243.8016470 & -23.3127069 & 7.1572953 & -8.8631422 & -26.3450177 & 6050388913137070848 & N & YNY \\
243.8311704 & -25.6701140 & 7.4134597 & -9.1980374 & -27.1873024 & 6048740710844628864 & N & YNN \\
243.8642421 & -22.6577711 & 7.1951507 & -10.8897003 & -24.8620069 & 6242599763466335104 & N & YNY \\
\hline
\end{tabular}
\caption{First 5 entries of the combined sample (532 members) described in Sect.~\ref{sect_common}. The sixth column
is \gaia DR2 \texttt{source\_id}. The entire table is provided in the electronic version of this article. The control column
indicates if a target is part of the control sample following a Yes/No nomenclature. All the coordinates are given in J2015.5
epoch (as in \gaia DR2).}
\label{tab:tb_3}
\end{center}
\end{table*}

The common sample contains 243 potential Ophiuchus member candidates that were not included in the three catalogues
used to construct the control sample described in Sect.~\ref{sect_sample_control}. In order to gather further information
about these sources, we queried the \simbad database using as identifier the \gaia DR2 \texttt{source\_id}. This query
returned 77 objects, that is, to date 166 objects in our common sample do not appear linked to any publication according
to the \simbad service. Combining the results obtained by  \cite{Kraus_2007}, \cite{2012ApJ...758...31L}, and 
\cite{2015MNRAS.448.2737R}, we find that 43 targets out of the 77 known objects have been associated with the USco
subgroup of the Scorpius-Centaurus OB association.

\section{Results}

\subsection{Astrometric properties}
\label{sect_astrometry}

\begin{figure}[ht!]
  \centering \includegraphics[width=\columnwidth, trim = 0 40 0 0]{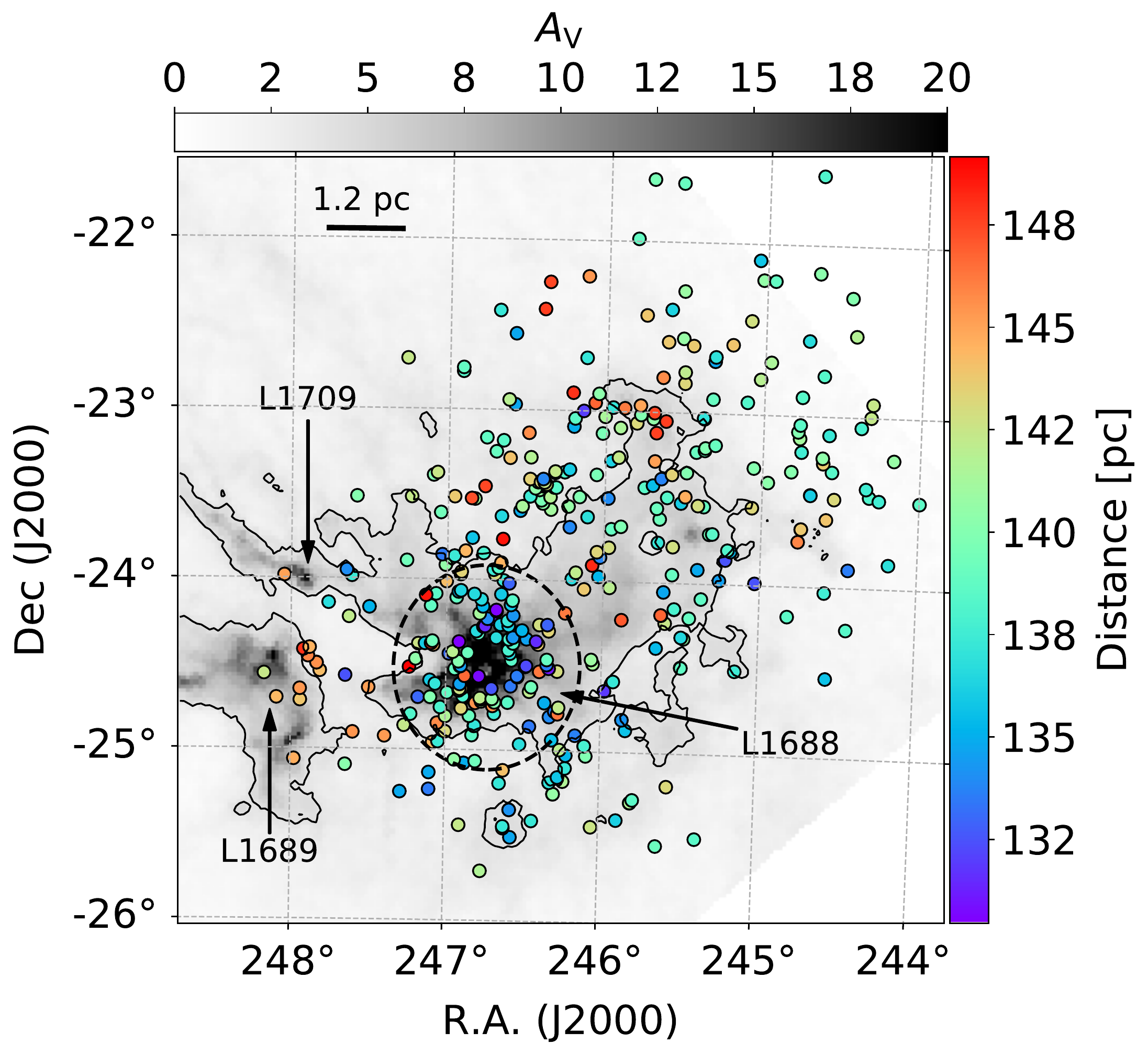}
  \caption{Common sample overlaid on top of an extinction map. The contour delimits the $A_\mathrm{v} = 4$
  regions. The arrows indicate the location of the main Lynds dark clouds in this area, while the black dashed line
  encompasses the $r \le 0.6\degr$ region centred on the extinction peak. The size scale is calibrated for the average
  distance to the common sample.}
   \label{fig:fig_7}
\end{figure}

The sky-projected distribution of the common sample is shown in Fig.~\ref{fig:fig_7} superimposed on the extinction map
produced by the COMPLETE project \citep{2006AJ....131.2921R} applying the NICER algorithm \citep{2001A&A...377.1023L}
to 2MASS observations. Nearly one-third of the common sample objects (125 sources) are found within a distance
$r \leq 0.6\degr$ from the extinction peak (roughly at $\alpha, \delta = 246.7\degr, -24.5\degr$) of the main dark
cloud of Ophiuchus, the Lynds 1688 dark cloud \citep[L1688; see e.g. ][]{2008hsf2.book..351W}. Most of these
objects (113) are also included in the control sample. The apparent lack of sources at the innermost core of this
cloud is most likely a consequence of the very high extinction exceeding $A_\mathrm{V} > 20$ at this location.
We do not find a significant correlation between the distances to each object (indicated by the rainbow colour bar)
and the sky-projected location of the sample. 

\begin{figure}[ht!]
  \centering \includegraphics[width=1.0\columnwidth, trim = 0 40 0 0]{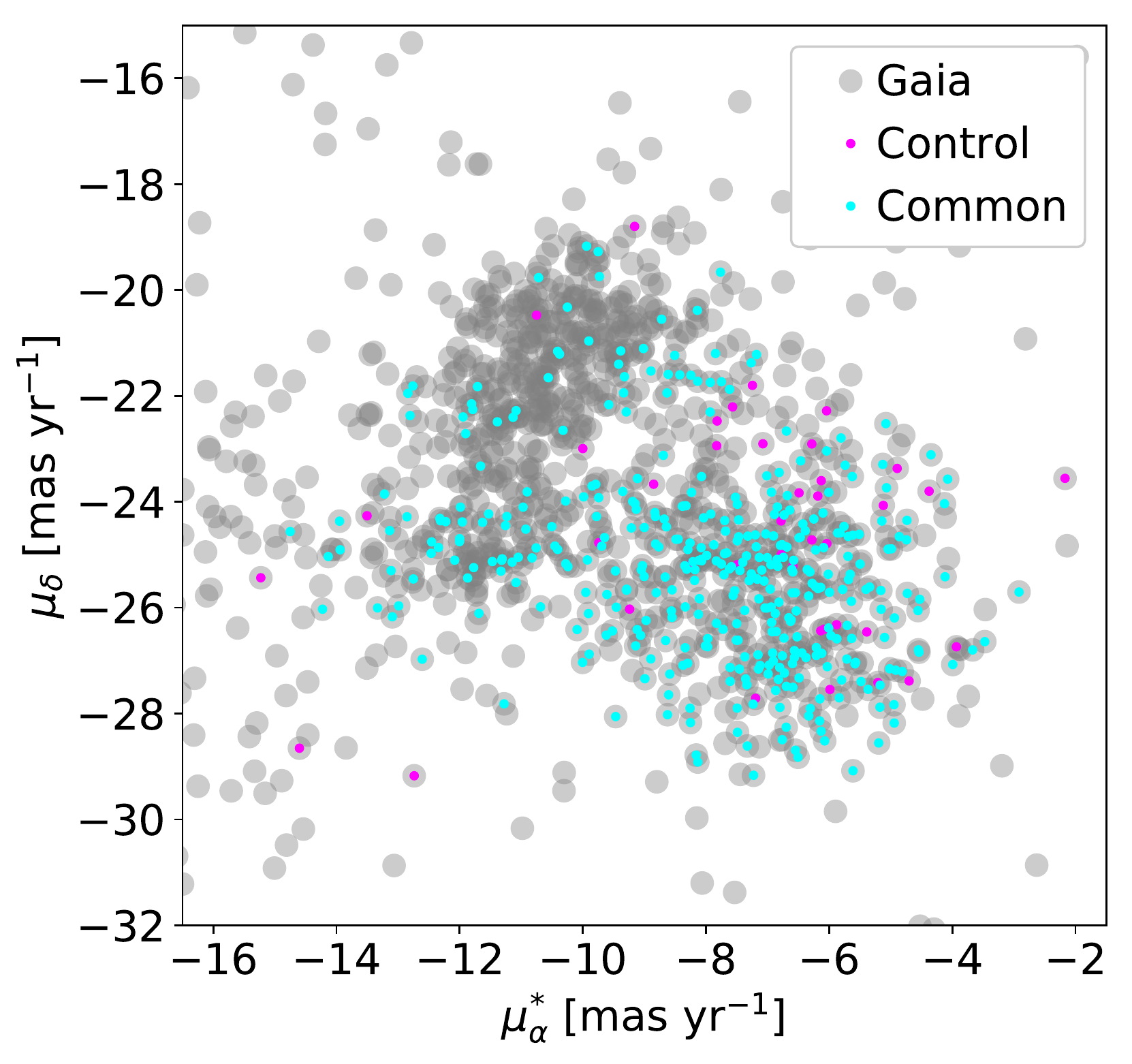}
  \caption{Same as Fig.~\ref{fig:fig_2}, including the common sample as cyan small circles. The 40 sources from
  the control sample not included in the common sample are shown as small magenta circles.}
  \label{fig:fig_8}
\end{figure}

\begin{figure}[ht!]
 \centering \includegraphics[width=\columnwidth, trim = 0 100 0 0]{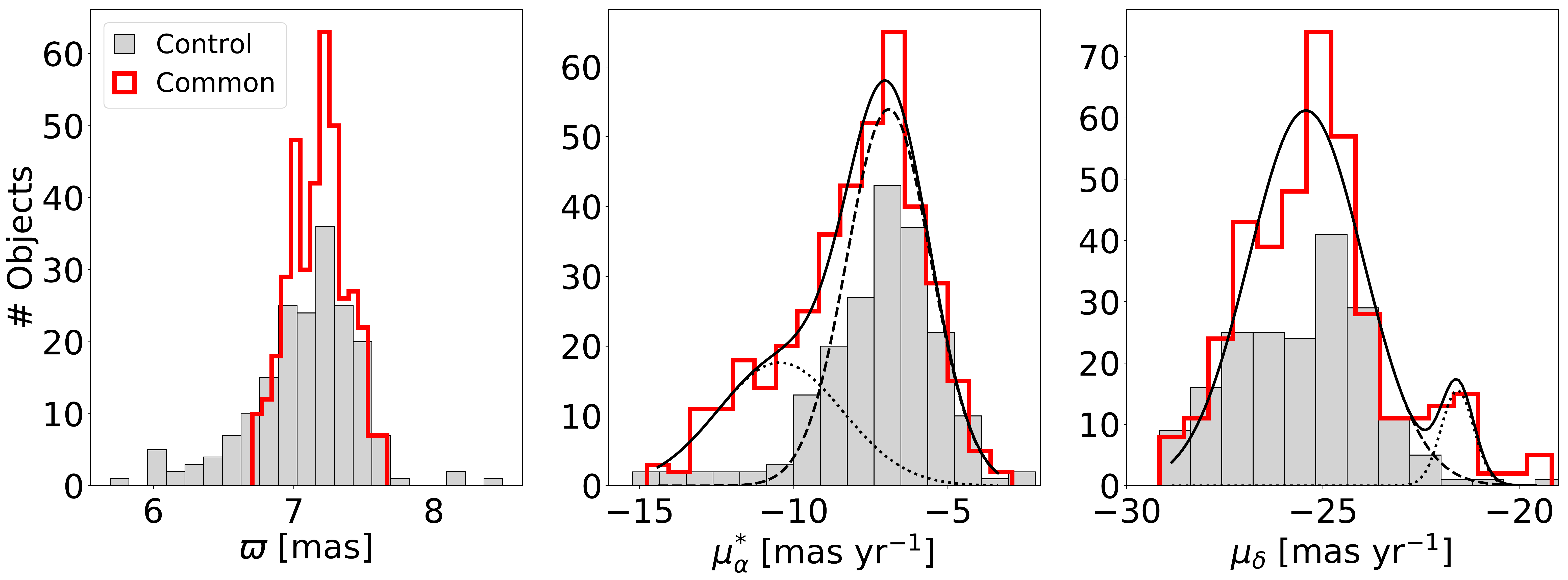}
  \caption{Histogram distributions of the common sample (see Sect.~\ref{sect_common}). A two-Gaussian fit is shown
  as a solid line superimposed on the proper motion histograms.}
  \label{fig:fig_9}
\end{figure}

Figure~\ref{fig:fig_8} shows the proper motions of the common sample (plotted as small cyan circles) overlaid on the
zoomed-in distributions of the \gaia sample (as grey circles). As in Fig.~\ref{fig:fig_2}, the diameter of the grey
circles equals the average error on \pmra of the \gaia sample. The 40 control sample sources not included in the
common sample are plotted as small magenta circles. Most of the common sample members are concentrated
around the apparent cluster located at higher \pmra values (Cluster 1 in Fig.~\ref{fig:fig_2}), but a non-negligible
fraction of these members are also found in the apparent cluster located towards lower \pmra and higher \pmdec
values (Cluster 2 in Fig.~\ref{fig:fig_2}). The parallax and proper motion histograms of the common sample are
shown in Fig~\ref{fig:fig_9}. The three distributions deviate from a single bell-shaped distribution and the parallax
histogram shows two separate peaks at $\varpi = 7.0$ and $\varpi = 7.2$. The parallax distribution is narrower than
the control sample counterpart (\para $\in [6.7:7.7]$ versus \para $\in [5.7:8.5]$). By computing the distance as the
inverse of the parallax, we obtain an average distance of $d = 139.4_{-3.8}^{+4.1}$ pc to the common sample.
The difference in parallax ranges means that the common sample members are distributed in a localised spatial
region with distances ranging from $[130:150]$ pc, whereas the control sample occupies a spatial region with
distances ranging between $[118:176]$ pc. The proper motion histograms show a secondary peak or shoulder
at \pmra$\sim-12$ and \pmdec$\sim-21$, as expected from Fig.~\ref{fig:fig_8}. We fitted a combination of two
Gaussian profiles to the \pmra and \pmdec distributions using a Levenberg-Marquardt algorithm; the fit did not
converge for the parallax distribution. The fits are shown as a solid (combined fit), dashed (main Gaussian), and
dotted (secondary Gaussian) lines overlaid on the histograms in Fig.~\ref{fig:fig_9}. The mean and full width at
half maximum (FWHM) of the Gaussian fits and the average astrometric properties of the common sample are
listed on Tables~\ref{tab:tb_4} and ~\ref{tab:tb_5}, respectively. 

It is likely that our common sample contains members of the USco region given the similarity between the
proper motions of USco and \rhooph, the proximity between these two regions, and the large extension of
USco across the sky \citep[e.g. ][]{Mamajek_2008, 2002AJ....124..404P, 2008hsf2.book..235P}. We
used the USco control sample with trigonometric parallaxes discussed by \cite{2018MNRAS.tmpL..38G}
to compare the astrometric properties of both samples. The average parallax and proper motions of the
USco sample (\para$=7.1\pm0.5$,  \pmra$=-11.7\pm3.1$, and \pmdec$=-23.9\pm1.9$) are indicated by
the blue vertical dot-dashed lines in Fig.~\ref{fig:fig_9}. The average \pmra of the USco sample coincides
with the secondary peak of the \pmra histogram distribution of our common sample, which we attribute to
the presence of USco stars in our sample.

\begin{table}
\begin{center}
\begin{tabular}{ccc}
\hline
 & \pmra & \pmdec \\
 & $[$mas yr$^{-1}]$ & $[$mas yr$^{-1}]$ \\
\hline 
\hline
Mean & $-6.9\pm0.1$ & $-25.4\pm0.1$ \\
FWHM & $3.2\pm0.1$ & $3.4\pm0.1$ \\
\hline 
Mean & $-10.4\pm0.5$ & $-21.6\pm0.1$ \\
FWHM & $4.8\pm0.8$ & $1.0\pm0.4$ \\
\hline
\end{tabular}
\caption{Mean and FWHM of the Gaussian fits shown in Fig.~\ref{fig:fig_9} for the main and secondary Gaussians
(upper and lower halves, respectively). The errors are the $1\sigma$ uncertainties of the fit.}
\label{tab:tb_4}
\end{center}
\end{table}

\begin{table}
\begin{center}
\begin{tabular}{cc}
\hline
 Dimension & Value\\
\hline
\hline
\ra &  $246.2^\circ\pm0.8$\\
\dec & $23.9^\circ\pm0.8$\\
\para & $7.2\pm0.2$ mas \\
\pmra & $-8.0\pm2.2$ mas\,yr$^{-1}$\\
\pmdec & $-25.2\pm1.9$ mas\,yr$^{-1}$\\
$X$ & $132.1\pm3.8$ pc\\
$Y$ & $-15.8\pm1.4$ pc\\
$Z$ & $41.9\pm2.6 $ pc\\
\hline
\end{tabular}
\caption{Mean and standard deviation ($1\sigma$) of the common sample across different dimensions. The Cartesian coordinates
are given in the Galactic reference frame ($X$, $Y$, and $Z$ point towards the Galactic centre, the direction of the Galactic rotation,
and the North Galactic Pole, respectively).}
\label{tab:tb_5}
\end{center}
\end{table}

\subsection{Colour magnitude diagram}
\label{sect_cmd}

\begin{figure}[ht!]
  \centering \includegraphics[width=1.0\columnwidth, trim = 0 30 0 0]{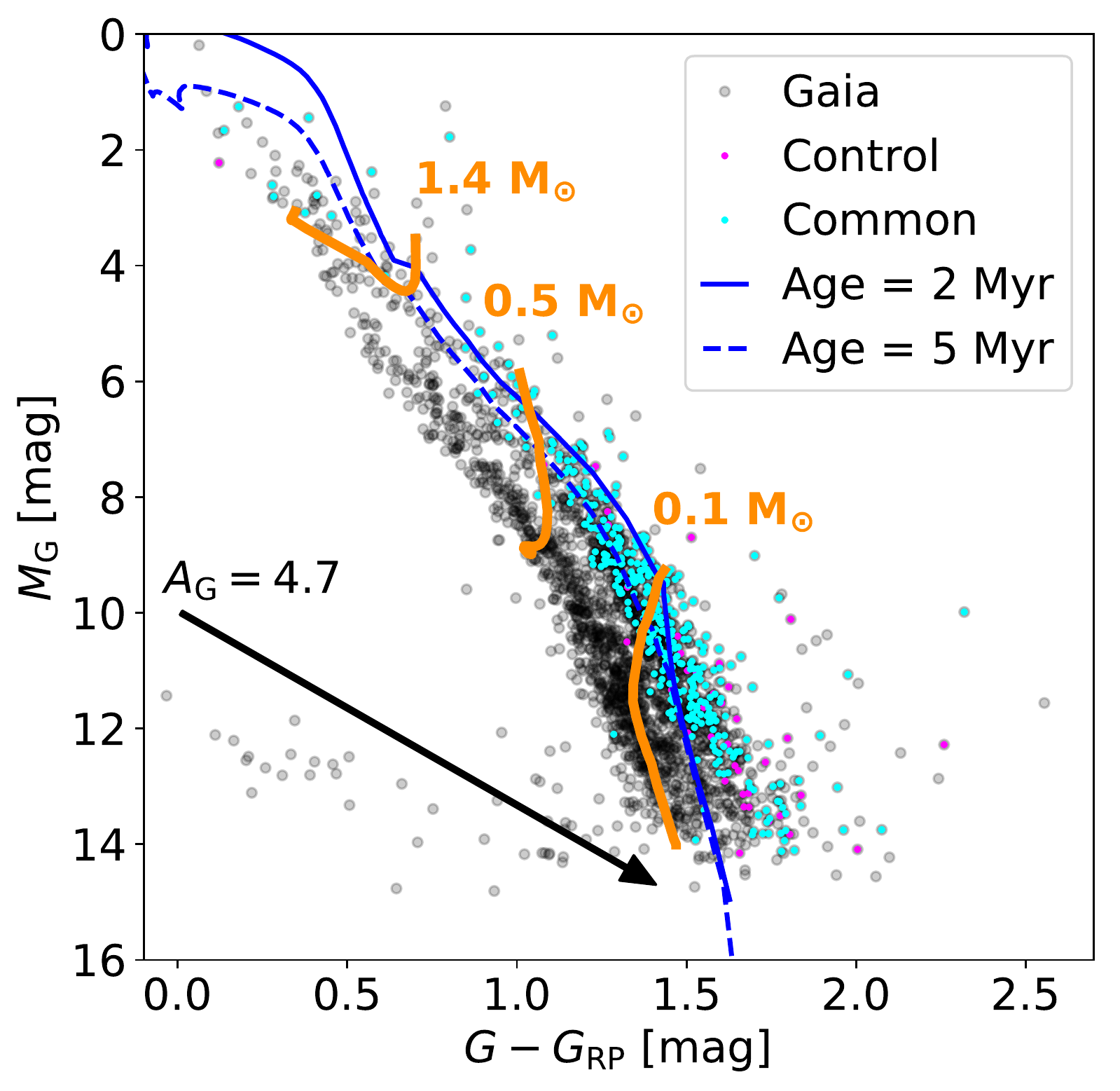}
  \caption{Colour magnitude diagram. The control sample sources not included in the common sample are shown as
  small magenta circles. The diameter of the black circles is larger than the average photometric error. The extinction
  vector is computed for the average \Ag derived from a subset of sources (see Sect.~\ref{sect_cmd}).}
  \label{fig:fig_10}
\end{figure}

Fundamental properties like stellar masses or ages can be estimated from a colour magnitude diagram (CMD),
if the extinction for each object is known. There are photometric measurements for all the objects in our common
sample in the three \gaia bands \citep [broad $G$ and the two narrower $G_\mathrm{BP}$ and $G_\mathrm{RP}$;][]{2018}
bands, but the extinction is estimated for only 95 of these objects and has an average value of \Ag$= 1.7\pm0.6$. As noted by
\cite{2018A&A...616A...8A} the \gaia extinctions are not accurate at the individual star level. Most importantly for our
particular case, the effective temperatures and extinctions listed on the \gaia DR2 are based on a naked stellar model
that does not include the contribution from the dust in the SFR or the protoplanetary discs that
probably surround an important fraction of the common sample sources. Therefore we looked for published extinction
values of our sources. \cite{Dunham_2015} gave extinction values in the $V$ band, \Av, for 83 objects with an average
value of \Av$ = 6.0 \pm 4.0$. \cite{2015MNRAS.448.2737R} estimated \Av for another 20 objects with an average value of
\Av$= 0.9 \pm 0.6$. The large difference between both studies is because, by construction, the sample by
\cite{Dunham_2015} is composed of YSOs surrounded by discs, while the sample discussed by \cite{2015MNRAS.448.2737R}
is dominated by objects without detected discs. Combining both studies we obtain \Av
estimates for $26\%$ of the sources in the common sample that have an average extinction of \Av$= 5.0\pm4.1$. Using the
\gaia DR2 extinction coefficients listed in the filter profile service provided by the Spanish Virtual Observatory
\citep[SVO\footnote{https://svo.cab.inta-csic.es/main/index.php};][]{2012ivoa.rept.1015R} this average value translates
into \Ag$= 4.7$. As the individual extinctions for most of the sources remain unknown to date, we opted for constructing
an extinction uncorrected CMD. For this purpose we used the $G$ and $G_\mathrm{RP}$ bands, as their characteristic error
is lower than in the $G_\mathrm{BP}$ band; for our entire sample we obtain a $\delta G, \delta G_\mathrm{RP}$,
and $\delta G_\mathrm{BP}$ of $0.002, 0.007$, and $0.04$ mags, respectively. We used the individual parallaxes to
compute the absolute magnitudes for each source as $M = m + 5 (\log_{10} \varpi +1)$. We computed the extinction
vector using the previously mentioned average \Ag  to get a rough idea of the extinction effect. Figure~\ref{fig:fig_10}
shows the obtained diagram; the common sample is plotted as small cyan circles superimposed on the grey circles
that represent the \gaia sample. As in Fig.~\ref{fig:fig_8}, the 40 sources that belong to the control sample but not included
in the common sample are shown as magenta small circles. Except for a few objects, the common sample occupies a
separate region in this plot, as expected if this sample is significantly younger than the rest of the stars in the \gaia
sample. Given the high extinction of the \rhooph region the common sample is most likely composed of objects located
in the near side of the cloud surface, as those in the inner regions of the cloud are probably too obscured to be detected by
\gaia. Median ages for \rhooph members at the cloud surface range from 2 Myr to 5 Myr \citep{2008hsf2.book..351W}. To put
our  sample in context we gathered the BT-Settl/CIFIST \citep{2015A&A...577A..42B} and Parsec \citep{Marigo_2017}
stellar evolutionary models for stellar masses below and above $1.4\,M_{\odot}$, respectively.
The 2 and 5 Myr isochrones
drawn from these models are shown as a solid and a dashed blue line overlaid on the CMD, respectively, while the
evolutionary tracks for three different stellar masses ($0.1, 0.5$, and $1.4\,M_{\odot}$) are indicated by orange thick
lines. Most of the common sample objects are located above the isochrone. Similarly, the bulk of the common sample
($\sim 90\%$ of it) lies below the $0.5\,M_{\sun}$ evolutionary track.

\subsection{Potential discs}
\label{sect_discs}

Given the youth of \rhooph, a large fraction of our common sample members might be surrounded by protoplanetary discs.
The presence of such a disc, as well as its evolutionary stage, can be inferred from the warm dust emission at infrared
wavelengths. We used the crossmatched
tables provided by the \gaia consortium (available in the \gaia archive) \texttt{gaiadr1.tmass\_original\_valid} and
\texttt{gaiadr2.tmass\_best\_neighbour} to obtain the near-infrared photometry for the common sample
sources. By combining these two tables we retrieved the 2MASS photometry for 382 sources.
We then repeated the same procedure but using the \texttt{gaiadr2.allwise\_best\_neighbour} and \texttt{gaiadr1.allwise\_original\_valid}
tables to obtain the mid-infrared photometry measured by the Wide-Field Infrared Survey Explorer \citep[WISE;][]{Wright_2010},
retrieving observations for 332 sources. All the retrieved 2MASS photometry has quality flag "A" at the three bands, but this
is not the case for the WISE photometry. Therefore, before combining these outputs we disregarded the objects with
photometric quality flags different than "A" or "B" at any of the WISE bands. Furthermore, we only considered objects
with contamination and confusion flags (\texttt{cff}) equal to "0" in all WISE bands, and with extended source flags (\texttt{ex})
lower than 2. At this stage, we inspected by eye the WISE images of every source to remove those affected by artefacts.
The WISE band 4 photometry can be strongly affected by background emission, which might be mistakenly associated with
the warm dust emission produced by a circumstellar disc \citep{2012MNRAS.426...91K}. To minimise this potential
confusion we used the \texttt{Astropy/photutils} package to perform aperture photometry to measure the signal to noise
at the peak of the WISE band 4 image of each source. We used an aperture mask with radius $r_\mathrm{ap}=12\arcsec$
to measure the peak emission and a sky aperture ring of $r_\mathrm{in} = 20\arcsec$ and $r_\mathrm{out} = 25\arcsec$
to estimate the background signal; both apertures were centred on the \gaia DR2 coordinates (\ra and \dec) for each source.
We filtered out all the sources with peak emission S/N $\leq 4$. Applying this selection criteria we finally obtained a sample
of 48 sources,  of which 25 belong to the control sample, with 2MASS and WISE photometry.

We combined the \gaia photometry with the infrared photometry to build up the spectral energy distributions (SEDs) for the 48
sources previously selected. To convert from magnitudes to flux units (Jy) we retrieved the effective wavelengths and zero points
(ZP) for each photometric band from the SVO. The total uncertainty of the fluxes was computed as the quadratic sum of the errors
associated with the photometric extraction process,  ZP, and systematic errors. The \gaia photometric uncertainties are
negligible compared to those from the 2MASS and WISE. The largest source of uncertainties are the systematic errors associated
with the WISE photometry, which is tied to the Spitzer photometric calibration \citep{Jarrett_2011}. In this work we adopt an absolute
calibration uncertainty of 5\% for the W1, W2, and W3 bands, and 10\% for the W4 band. Given the broad wavelength coverage
of the WISE filters, a colour correction must be applied when transforming magnitudes to fluxes. For each source we computed
the $\alpha$ flux slope ($F_\nu \sim \nu^\alpha$) for the different WISE bands and then applied the correction factors following
Section VI of the Explanatory Supplement to the WISE All-Sky Data Release Products\footnote{http://wise2.ipac.caltech.edu/docs/release/allsky/expsup/sec4\_4h.html}.

\begin{figure}[ht!]
  \centering \includegraphics[width=1.0\columnwidth, trim = 0 30 0 0]{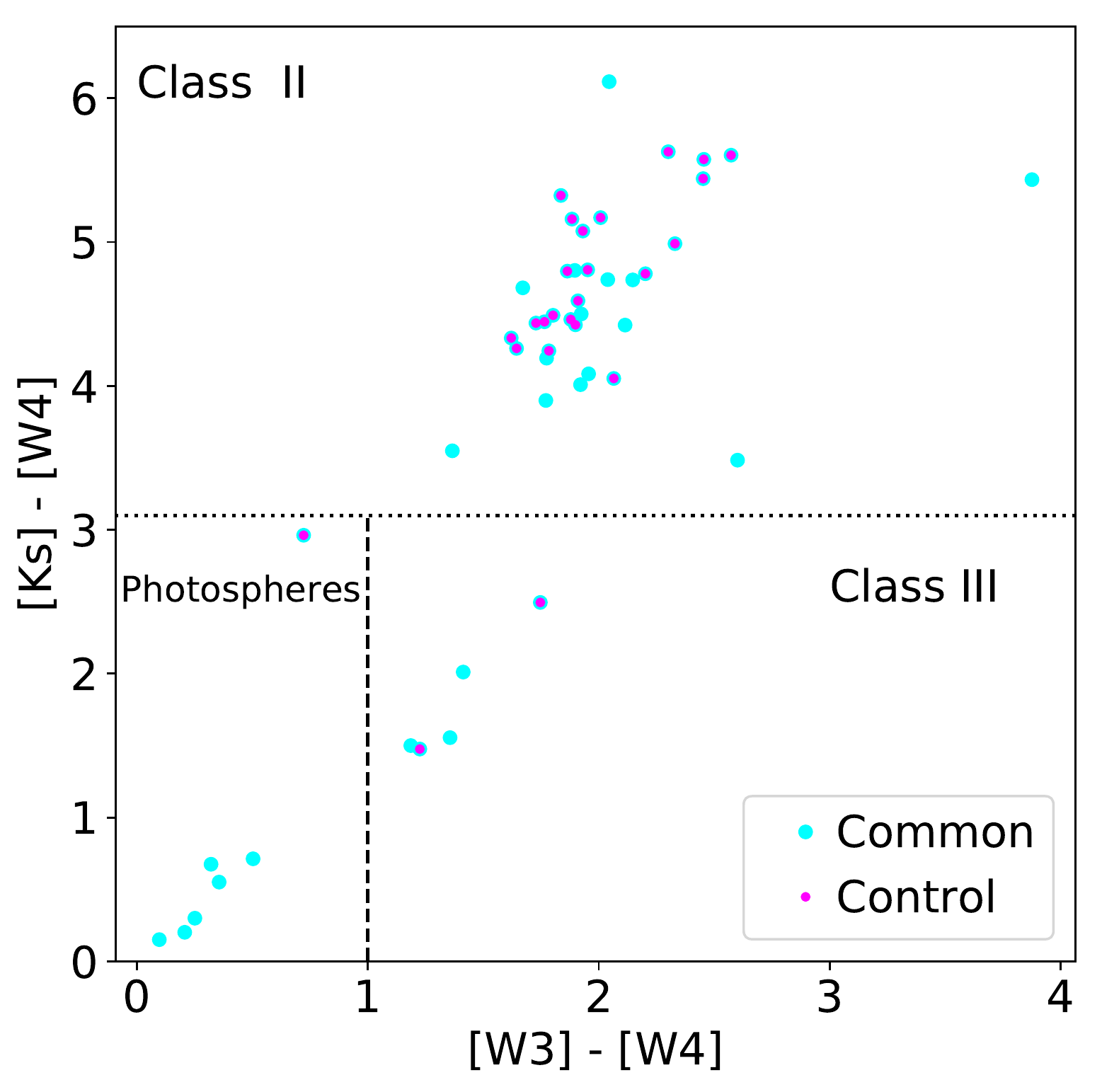}
  \caption{Colour-colour plot of the common sample with high quality infrared photometry. The dotted line indicates
  the traditional boundary between the Class II and III stages. The vertical dashed line separates bare photospheres
  from Class III discs.}
  \label{fig:fig_11}
\end{figure}

Traditionally, the protoplanetary discs have been classified according to the infrared slope of their SED defined as 
\begin{equation}
 \alpha_\mathrm{IR} = 
  {\frac {log(\lambda_1 F_{\lambda_{1}})  -  log(\lambda_0 F_{\lambda_{0}})}     {log(\lambda _{1})-  log(\lambda _{0})}},
 \end{equation} 

where $\lambda_1$ and $\lambda_0$ correspond to $\sim$2 and $\sim$22 $\microm$ \citep{1984ApJ...287..610L, 
1987IAUS..115....1L}. This way, objects with \alphaIR$>0.3$ (i.e. those with an increasing SED towards the mid-infrared)
are labelled as Class I and are associated with YSOs surrounded by a primordial envelope and massive,
little evolved disc. Objects with $-0.3>$\alphaIR$>-1.6$ are classified as Class II sources and represent a more evolved
stage during the disc evolution, while those with $0.3>$\alphaIR$>-0.3$ are classified as flat SED sources and show
intermediate properties between the Class I and II stages \citep{1994ApJ...434..614G}. Objects with \alphaIR$<-1.6$
are classified as Class III and they are associated with tenuous discs. Applying this criteria and using the $K_\mathrm{s}$
($\lambda_\mathrm{eff} \sim 2.2\microm$) and W4 ($\lambda_\mathrm{eff} \sim 22.1\microm$) bands, we find 36
Class II and 12 Class III objects within the 48 objects with infrared photometry contained in the common sample. This
traditional classification is useful and allows for comparison with previous works, but it has some limitations \citep[see e.g.]
[]{2007ApJ...667..308C, 2010ApJ...718.1200M}. In particular, as already noted by \cite{2009ApJS..181..321E}, the
Class III encompasses two different types of objects: those surrounded by a tenuous disc, and those with no detectable
infrared excess (i.e. bare photospheres). Inspecting the computed SEDs we find several of these objects within the Class III
of our sample. Therefore we applied an extra classification based on the infrared slope between $\sim12\microm$ and
$\sim22\microm$ ($\alpha_\mathrm{W34}$). We classify as bare photospheres the Class III objects with $\alpha_\mathrm{W34} < -1.6$,
and reclassify the Class III objects as those with \alphaIR$<-1.6$  and $\alpha_\mathrm{W34} > -1.6$. Using this classification
scheme we identify 7 bare photospheres and 5 Class III sources, as reflected in Fig.~\ref{fig:fig_11}. Therefore, we find
a disc fraction of $\sim85\%$ in this subsample of 48 objects. Bearing in mind the strong selection biases of our sample,
we note that this high disc fraction is consistent with an age of $\sim2$ Myr according to different studies \citep[e.g.][]
{2015A&A...576A..52R, 2010A&amp;A...510A..72F, 2009AIPC.1158....3M}. A table with the photometry for these 48 objects
is given in the Appendix (Table~\ref{tab:ap_tb_4}) and is available in the electronic version of this article. This
table contains a ``status'' keyword that indicates if a disc is included in the control sample (labelled as control), or if the
disc has currently no references in the \simbad service (labelled as new).  We find 12 new discs, 11 of which have
Class II SEDs and 1 of which have Class III SED (\gaia DR2 $\#$6051732000945974912). The SEDs for these 12 new
discs are shown in Fig.~\ref{fig:fig_12}, and their corresponding WISE W4 images are shown in the Appendix (Fig.~\ref{fig:ap_5}).
We note that the disc around the \gaia DR2 $\#6050899361406000000$ source shows a strong increase towards
22$\microm$ characteristic of a disc with a large dust depleted inner cavity as in for example  Sz\,91 \citep{2015ApJ...805...21C,
2016MNRAS.458L..29C, 2014ApJ...783...90T}.

\begin{figure*}[ht!]
  \centering \includegraphics[width=1.0\textwidth, trim = 0 70 0 0]{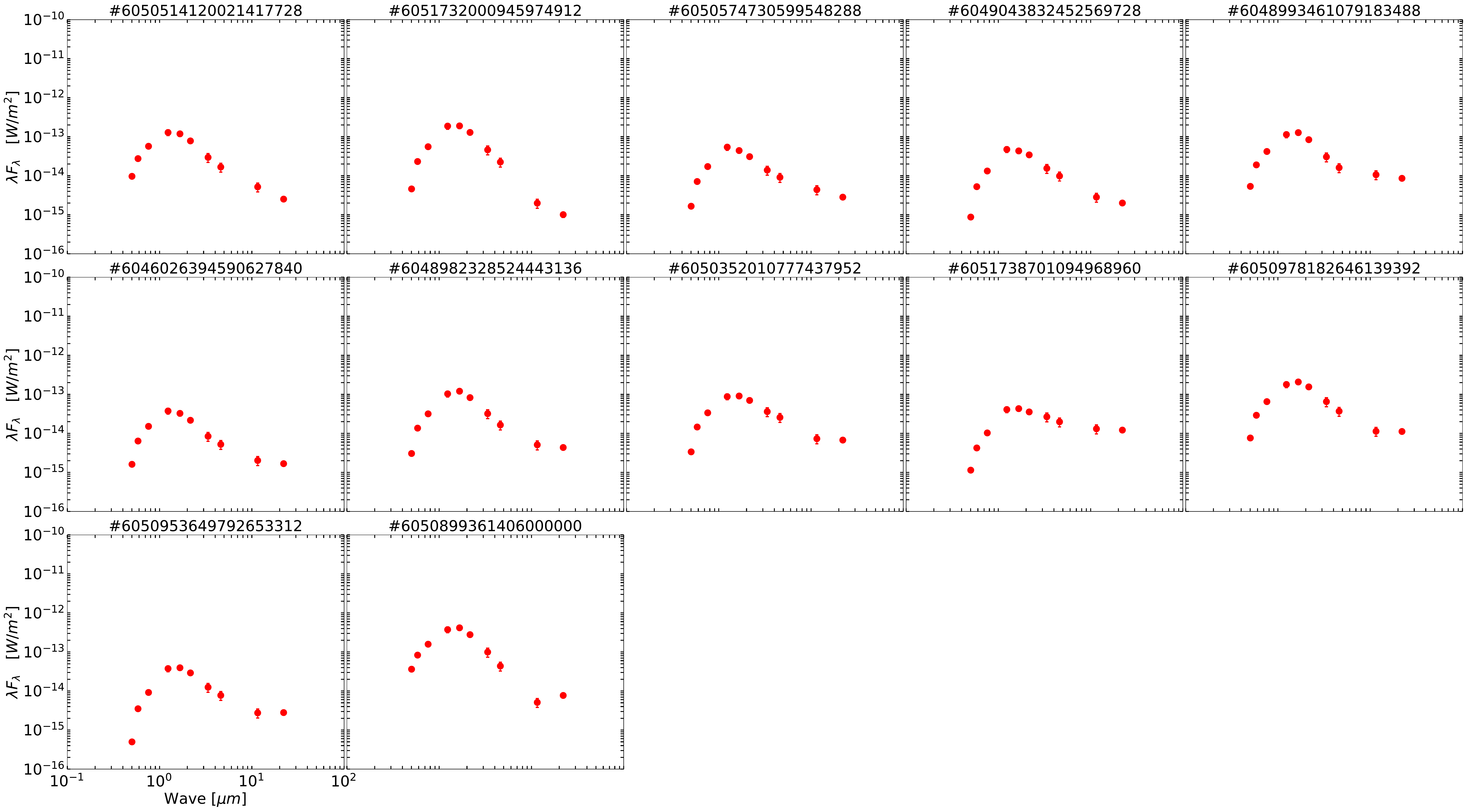}
  \caption{Spectral energy distributions of the 12 new discs detected. The title indicates their \gaia DR2 source ID. The
  panels are ordered by increasing $\alpha_{W34}$ slope. Object $\#6051732000945974912$ is the only Class III disc
  of this subsample, as the rest are Class II discs. Source $\#6050899361406000000$ shows the characteristic SED
  shape of a disc with a dust depleted inner cavity.}
  \label{fig:fig_12}
\end{figure*}

\section{Summary and conclusions}
\label{sect_conclusions}

In this work we have used three density-based clustering algorithms (\dbscan, \optics, and \hdbscan) to identify
potential new members of the \rhooph region in the \gaia DR2 catalogue. The members of the same SFR have
similar proper motions and spatial distributions, so these ML algorithms are well suited to detect
new member candidates.

We began by constructing a comprehensive sample of Ophiuchus members previously identified by
\cite{2008hsf2.book..351W}, \cite{2011AJ....142..140E}, and \cite{Dunham_2015}. Table.~\ref{tab:ap_tb_1} lists
the 465 elements of this catalogue. We then used the 2MASS source IDs of these objects to cross-match them
against the \gaia DR2 catalogue. We obtained a control sample composed of 188 elements by applying the
quality selection criteria described in Sect.~\ref{sect_sample_control} and removing the outliers. This catalogue
of bona fide members, which may be useful for future studies of the \rhooph cloud, is accessible via
Table.~\ref{tab:ap_tb_1} through the Control keyword. We then downloaded the \gaia DR2 sources contained
within a circular area of $r = 3.5\degr$ centred on \rhooph and with $5<$\para$< 9$, a region large enough to
encompass the entire control sample (Fig.~\ref{fig:fig_1}). We imposed the quality selection criteria outlined in
Sects.~\ref{sect_sample_control} and ~\ref{sect_sample_gaia}, and then separately applied the \dbscan, \optics,
and \hdbscan algorithms to the five astrometric dimensions ($XYZ$ Cartesian coordinates and the proper
motions \pmra and \pmdec) of this sample. The \dbscan code systematically finds one large cluster consistent
with the control sample astrometric properties, while \optics and \hdbscan identify two clusters, the largest of
which has astrometric properties similar to those of the control sample (Fig.~\ref{fig:fig_6}). Discussing in detail
the properties of the secondary cluster is out of the scope of this paper, and we defer its analysis to a separate
study. The size of the main cluster depends on the algorithm hyperparameters, ranging from 427 to 552 elements.
On average, the algorithms recover $\sim84\%$ of the control sample as shown in Table~\ref{tab:tb_2}. By
combining the different algorithm outcomes we constructed the minimum common sample. This sample,
presented in Table~\ref{tab:tb_3}, contains 391 candidate members of \rhooph including 148 control sample
elements and 166 sources that, to date, have no reference in the literature according to \simbad. Seventy-seven
sources appear associated either with \rhooph or USco \citep[i.e. ][]{2007ApJ...667..308C, Cheetham_2015,
2012ApJ...758...31L, 2015MNRAS.448.2737R}. The average distance to the common sample
($d = 139.4_{-3.8}^{+4.1}$ pc) is consistent, within error bars, with previous distance estimates towards the \rhooph
cloud \citep{2007ApJS..169..105M, Mamajek_2008, 2017ApJ...834..141O, 2018ApJ...869L..33O}. 

In Sect.~\ref{sect_astrometry} we analysed the astrometric properties of the common sample. The parallax distribution
is narrower than that of the control sample, ranging from $[6.7:7.7]$ and showing a double peak at \para$=7.0, 7.2$.
The proper motion histograms show double Gaussian-like profiles with main and secondary peaks at $-6.9\pm0.1$
and  $-10.4\pm0.5$ in \pmra, and $-25.4\pm0.1$ and $-21.6\pm0.1$ in \pmdec (Figs.~\ref{fig:fig_8} and ~\ref{fig:fig_9},
and Table~\ref{tab:tb_4}). The extinction uncorrected CMD is shown in Fig.~\ref{fig:fig_10} and discussed in Sect.~\ref{sect_cmd}.
The CMD suggests that the common sample is dominated by objects younger than 5 Myr with masses below $0.5\,M_\sun$,
which agrees with previous studies of this cloud \citep{2005AJ....130.1733W, 2011AJ....142..140E}, but we are cautious about
our conclusions given the large uncertainty associated with the (yet unknown) extinction of these sources. Finally, in
Sect.~\ref{sect_discs} we constructed the SEDs from optical to mid-infrared wavelengths for the subset of 48 objects with
high quality photometry in the \gaia, 2MASS, and WISE bands. We identified 36 Class II, 5 Class III objects, and 7 bare
photospheres by applying a colour selection criteria (Fig.~\ref{fig:fig_11}). Their photometry, along with their SED Class
and a flag indicating if the object has no references in SIMBAD, is given in Table.~\ref{tab:ap_tb_4}. Twelve of these objects
have no references on the \simbad service and show strong infrared emission, and their SEDs are shown in Fig.~\ref{fig:fig_12}.
Their corresponding WISE W4 maps are presented in Fig~\ref{fig:ap_5}.

This work reflects the potential of applying modern ML techniques to identify young stellar populations
in huge astronomical catalogues such as the \gaia DR2. Our results show that the \dbscan, \optics, and \hdbscan
algorithms identify a similar population of potential members of the \rhooph SFR. All of these algorithms recover
roughly 84\% of the control sample and the number of elements of the main cluster found does not present strong
variations when comparing the different algorithm outputs. The lack of a complete census of bonafide members of
\rhooph prevents us from performing a quantitative analysis to decide which algorithm is the more adequate to find
potential member candidates in this cloud. We therefore choose to focus our analysis in the minimum-sized common
sample composed of the cluster elements simultaneously found by the three algorithms. This common sample contains
fewer elements than the clusters identified by each method separately (391 members versus, e.g. 427 for \hdbscan with
mPts = 21; see Table~\ref{tab:tb_2}), so it is possible that because of our rather conservative approach we are
missing potential cluster members. Nevertheless, the entire list of candidate members found by the three algorithms
is given in Table~\ref{tab:tb_3} with the hope  that this sample will be useful for future studies by the community.

We are cautious about the possible generalisation of our analysis to other SFRs or stellar associations. While some
regions have a very compact spatial distribution \citep[e.g. IC 348,][and references therein]{Ru_z_Rodr_guez_2018},
others are distributed across several parsecs \citep[e.g. USco, ][]{2018MNRAS.tmpL..38G, 2008hsf2.book..235P}, and
furthermore some present a rich and complex spatial substructure \citep[like Taurus,][]{Joncour_2018}. Therefore it is
possible that the agreement in the outputs produced by \dbscan, \optics, and \hdbscan when analysing  \rhooph is not
reached when analysing other regions with different geometry and dynamics.

These tools are especially useful when the region under study is affected by a high extinction that hinders a sample
selection from a CMD \citep[as in e.g.][]{2018ApJ...868...32G}. However, we are aware of a number of limitations of
our study. To begin with, none of the algorithms discussed in this work take into account the measurement uncertainties
and, as shown through our analysis, the outcomes are sensitive to the hyperparameters given by the user. We attempted
to overcome these issues by combining the results of the three algorithms and focussing our analysis on the minimum
size common sample of the multiple cases listed in Table~\ref{tab:tb_2}. Another caveat to consider is that SFRs are very
complex environments and a fraction of their components might not be located in a cloud core but in a stream or filament-like
spatial structure \citep{Andr__2010}. This is indeed the case for \rhooph; at least two prominent filaments seem
to departure from the L1688 dark cloud \citep{2006AJ....131.2921R}. This complex spatial distribution makes it more
difficult for the clustering algorithms to identify the cloud members, and that can explain why the applied algorithms
recovered $\sim84\%$   of the sample rather than the entire control sample. This can also explain the narrow range
in the parallax distribution of the common sample when compared to that of the control sample. A third important caveat
is that, given the proximity in terms of parallax and coordinates, it is possible that a fraction of the stars in our sample
was formed in the $\sim10$ Myr old USco region \citep{2012ApJ...746..154P, 2016MNRAS.461..794P}. The remarkable
overlap between the average \pmra of the USco sample discussed by \cite{2018MNRAS.tmpL..38G} and the secondary
peak found in the \pmra histogram of our common sample, together with the fact that 43 members of the common sample
have been associated with USco, is an indicative of this possible contamination. A potential way to discern the nature of
these sources would be to analyse the radial velocities, which we have not included in our study given the lack of measurements
in the \gaia DR2. Future data releases are expected to include more radial velocity measurements and therefore will
become ideal catalogues to repeat studies like this one. In the meantime, the brighter objects of the common sample
could be followed up with spectroscopy in order to determine their effective temperatures and extinctions, and therefore
to estimate their ages and masses; this would also help to constrain their parental region.

It is not surprising that none of the studied objects with infrared photometry belong to the earlier Class 0 and I stages. These
objects are very embedded and extinct, and therefore they are relatively difficult targets for \gaia. The 12 discs that we discovered
are good candidates to be directly imaged at optical and near-infrared wavelengths with high-contrast instruments
such as SPHERE \citep{2008SPIE.7014E..18B} at the Very Large Telescopes (VLT), and at submillimeter and longer wavelengths
with the ALMA observatory. Given the high fraction of circumstellar discs that we found in our subset of 48 targets ($\sim85\%$),
the objects discussed which lack infrared photometry are promising targets to search for discs with future infrared facilities such as
the James Webb Space Telescope (JWST) and proposed missions such as the Space Infrared Telescope for Cosmology and
Astrophysics (SPICA). 

\begin{acknowledgements}
        We thank the anonymous referee for his/her comments and suggestions.
        H.C.gratefully acknowledges support from the European Space Agency (ESA) Research Fellow programme. C.C. was supported by the 2018
        ESAC trainee programme. L.C. was supported by CONICYT-FONDECYT grant number 1171246.\\

        This publication makes use of data products from \\
        1) the Two Micron All  Sky Survey, which is a joint project of the University of Massachusetts and the Infrared Processing and Analysis
        Center/California Institute of Technology, funded by the National Aeronautics and Space Administration and the National Science Foundation. \\
        2) the Wide-field Infrared Survey Explorer, which is a joint project of the University of California, Los Angeles,
        and the Jet Propulsion Laboratory/California Institute of Technology, funded by the National Aeronautics and Space Administration.\\
        3) the European Space Agency (ESA) mission {\it Gaia} (\url{https://www.cosmos.esa.int/gaia}), processed by the {\it Gaia} Data Processing
        and Analysis Consortium (DPAC, \url{https://www.cosmos.esa.int/web/gaia/dpac/consortium}). Funding for the DPAC has been provided by 
        national institutions, in particular the institutions participating in the {\it Gaia} Multilateral Agreement.
        \\
        This research has made use of \\
        1) the NASA/ IPAC Infrared Science Archive, which is operated by the Jet Propulsion Laboratory, California Institute of Technology, under contract with the National
        Aeronautics and Space Administration;\\
        2) the VizieR catalogue access tool, CDS, Strasbourg, France; and\\
        3) Astropy, a community-developed core Python package for Astronomy \href{http://www.astropy.org}{Astropy Collaboration 2013}. \\
\end{acknowledgements}

\bibliographystyle{aa.bst}      
\bibliography{biblio.bib}          

\begin{appendix} 

\begin{figure}[ht!]
  \centering \includegraphics[width=\columnwidth, trim = 0 40 0 0]{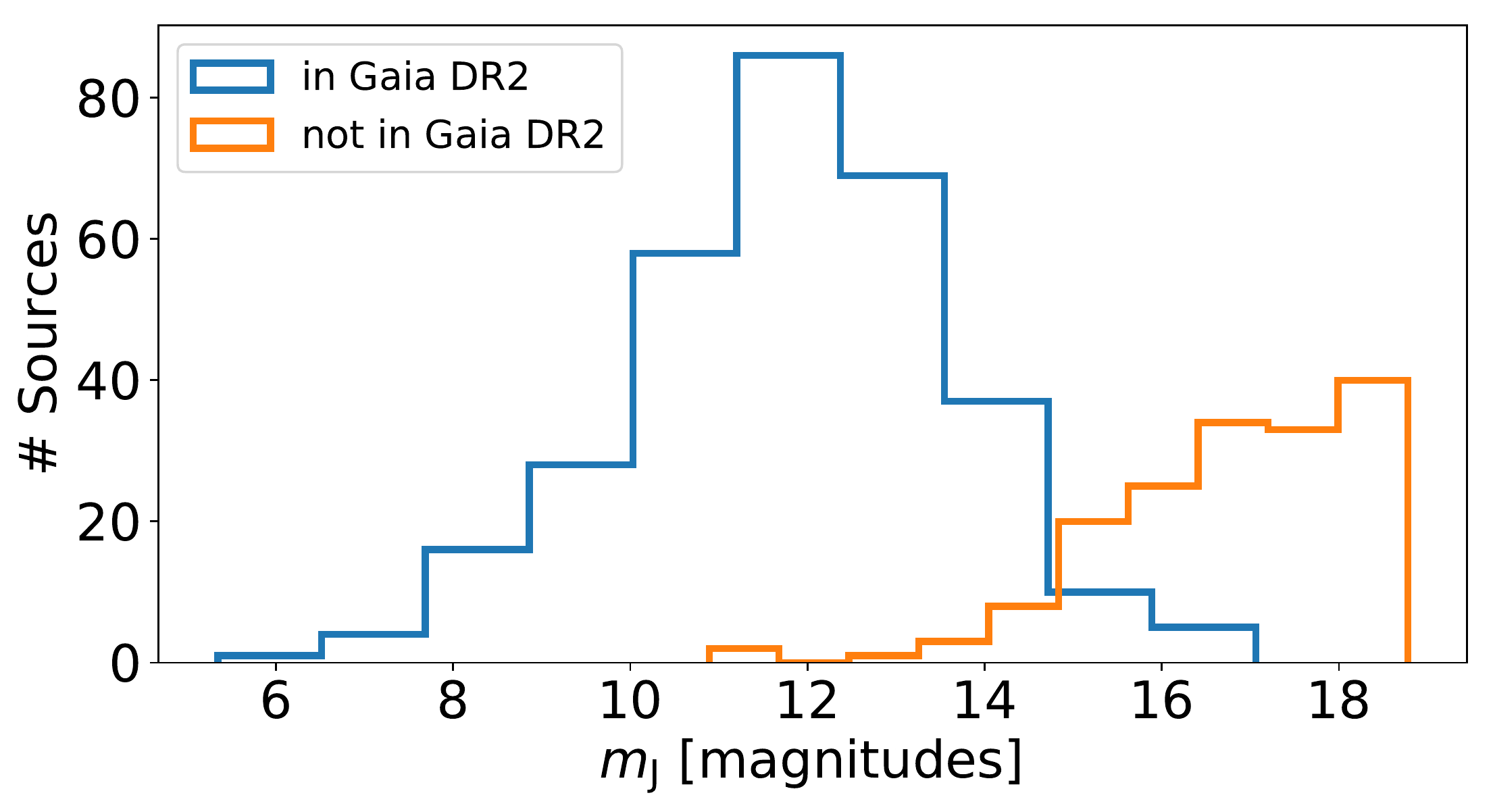}
  \caption{Apparent $J$ band magnitude histograms of the initial sample.}
   \label{fig:ap_1}
\end{figure}
\begin{figure}[ht!]
  \centering \includegraphics[width=\columnwidth, trim = 0 100 0 0]{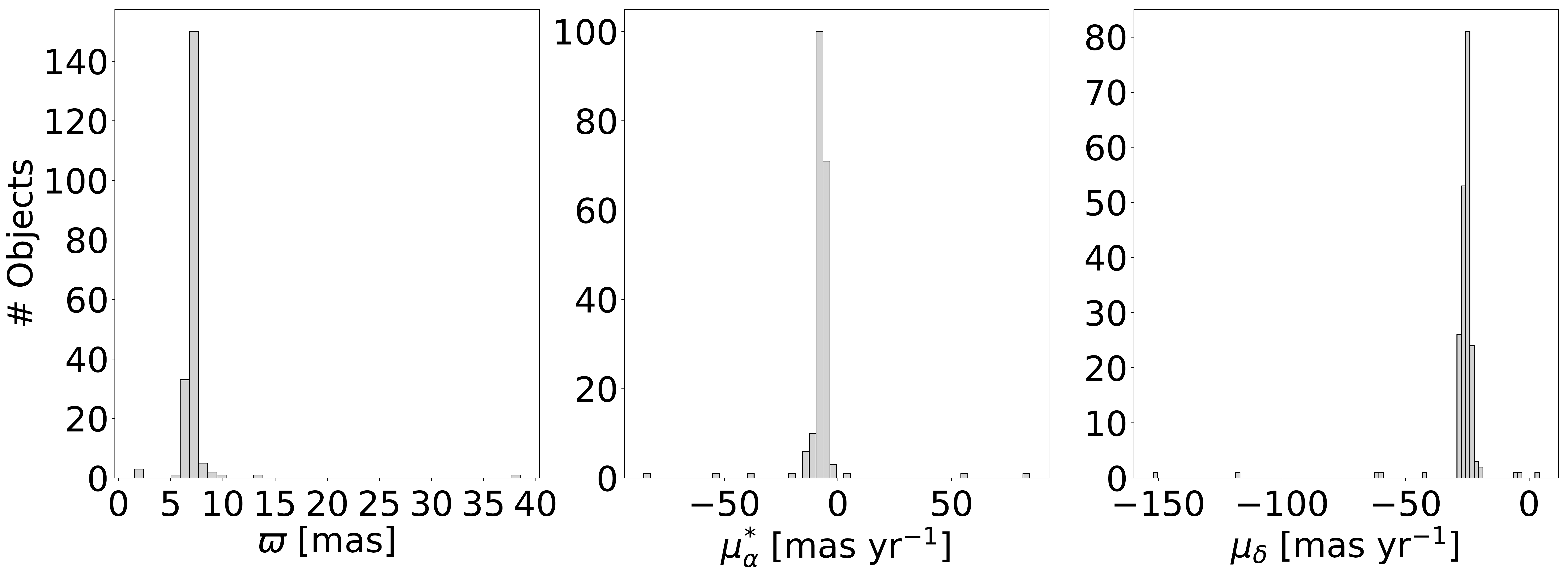}
  \caption{Parallax and proper motion histograms of the astrometrically cleaned sample. We note the few outliers located far away from
  the histogram peaks.}
   \label{fig:ap_2}
\end{figure}
\begin{figure}[ht!]
  \centering \includegraphics[width=\columnwidth, trim = 0 100 0 0]{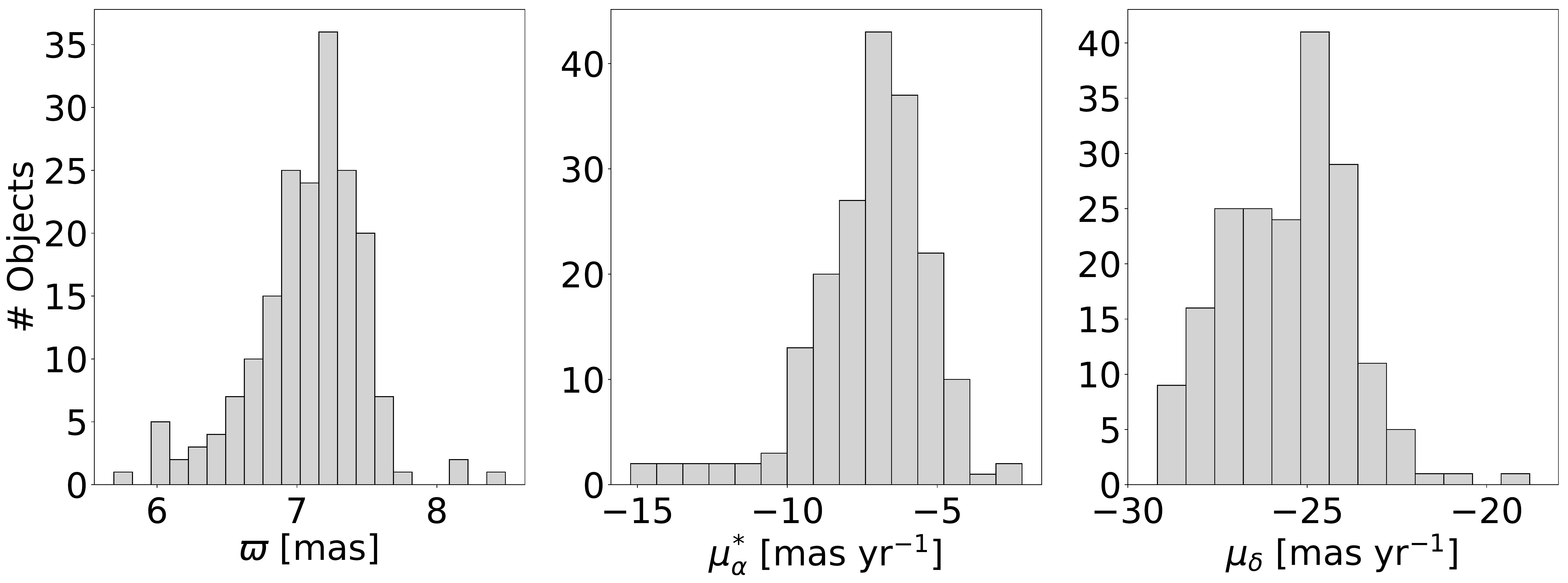}
  \caption{Same as Fig.~\ref{fig:ap_2} but for the control sample (obtained after removing the outliers listed
  in Table~\ref{tab:ap_tb_2} from the astrometrically cleaned sample).}
   \label{fig:ap_3}
\end{figure}
\begin{table}[ht!]
\begin{center}
\begin{tabular}{cccc}
\hline
2MASS & Refs. & Control &       DR2 Source ID\\
\hline
\hline
16261949-2437275 & 1, 2, 3 & Y & 6049122310095142656 \\
16262083-2428395 & 1, 3 & N &  \\
16262096-2408468 & 3 & Y & 6049357429490158336 \\
16262097-2408518 & 1, 2 & Y & 6049357433785068672 \\
16262138-2423040 & 1, 3 & N &  \\
\hline
\end{tabular}
\caption{Extract of the initial sample presented in Sect.~\ref{sect_sample_initial}. The entire table is listed in the electronic
version of the article. References are coded as  1: \cite{2008hsf2.book..351W}, 2: \cite{2011AJ....142..140E}, and 3: \cite{Dunham_2015}.
Note: Five targets have duplicate \gaia source\_id: 2MASS J16222099-2304025, 2MASS J16233609-2402209, 2MASS J16253958-2426349, 
2MASS J16275565-2444509, and 2MASS J16282373-2441412. These duplicates are removed when applying the selection
criteria described in Sect.~\ref{sect_sample_control}, but we keep them in the table for consistency.}
\label{tab:ap_tb_1}
\end{center}
\end{table}
\begin{table}[ht!]
\begin{center}
\begin{tabular}{cccc}
\hline
2MASS ID & \para & \pmra & \pmdec \\
 & [$\mathrm{mas}$] & [mas yr$^{-1}$] & [mas yr$^{-1}$] \\
\hline
\hline
16293441-2452292 & 8.9451 & -37.9989 & -118.5701 \\
16442882-2412242 & 2.1726 & -1.8833 & -5.9312 \\
16245652-2459381 & 1.4977 & -7.6758 & -3.7136 \\
16265752-2446060 & 10.2548 & 56.4965 & -59.1305 \\
16282516-2445009 & 13.1651 & -52.7356 & -62.1384 \\
16244802-2440051 & 38.5067 & -85.4146 & -152.0051 \\
16260302-2423360 & 7.4456 & -20.1836 & -26.7646 \\
16273052-2432347 & 9.3827 & 84.3542 & -41.9827 \\
16273714-2359330 & 2.0953 & 4.7860 & 4.0732 \\
\hline
\end{tabular}
\caption{Objects excluded from the control sample. They appear as outliers on the bell-shaped histograms shown in Fig.~\ref{fig:ap_2}.}
\label{tab:ap_tb_2}
\end{center}
\end{table}

\begin{figure}[ht!]
  \centering \includegraphics[width=\columnwidth, trim = 0 100 0 0]{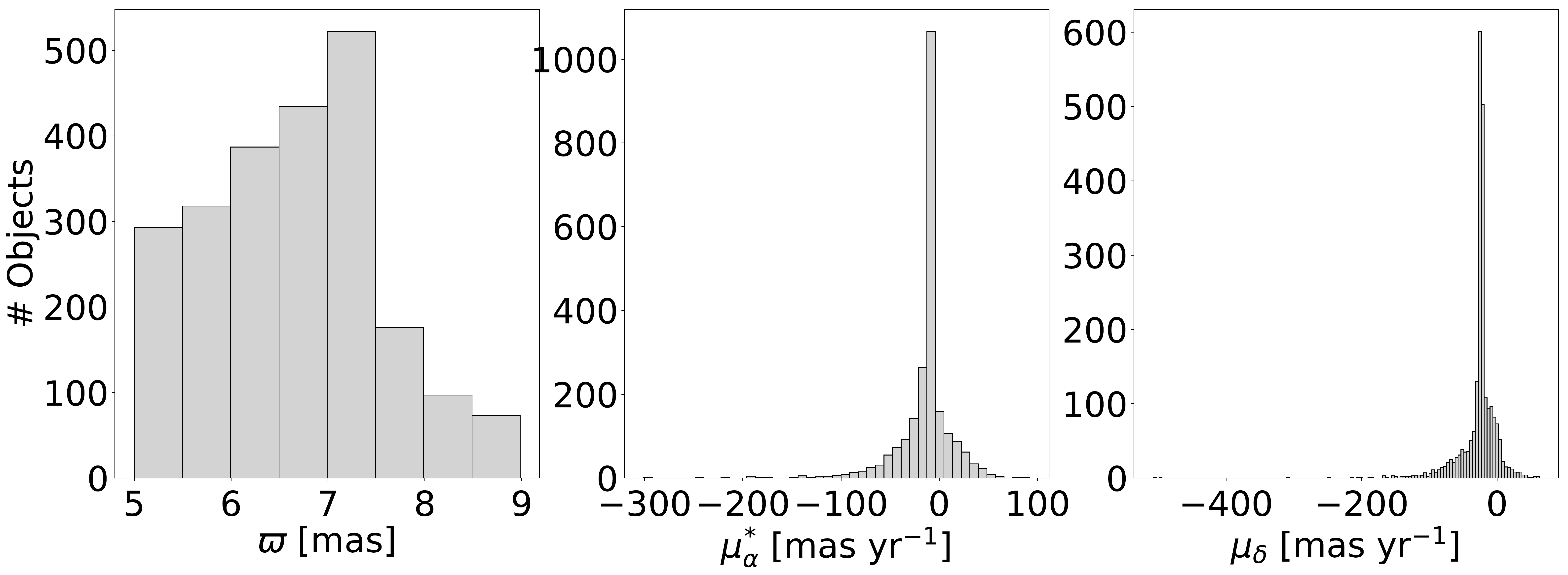}
  \caption{Same as Fig.~\ref{fig:ap_2} but for the \gaia cleaned sample.}
   \label{fig:ap_4}
\end{figure}
\begin{table}[ht!]
\begin{center}
\begin{tabular}{cccc}
\hline
\hline
DR2 Source ID& Ref. & \Av & \Ag  \\
 &  & $\mathrm{mag}$ & $\mathrm{mag}$ \\
\hline
6046062364938324608 & 1 & 4.2 & 0.8 \\
6047590243721186688 & 1 & 4.4 & 2.6 \\
6049067776894321024 & 1 & 3.9 & 1.3 \\
6049094616145974272 & 1 & 6.3 & 2.2 \\
6049095406419962112 & 1 & 7.9 & 1.9 \\
6049101866050768384 & 1 & 2.9 & 1.9 \\
6049118388788178816 & 1 & 4.0 & 2.8 \\
6049122310095142656 & 1 & 1.6 & 2.2 \\
6049142410542091648 & 1 & 4.5 & 2.8 \\
6049161514556593920 & 1 & 15.9 & 2.0 \\
6049162545348745984 & 1 & 10.8 & 2.1 \\
6049331526542488064 & 1 & 6.6 & 2.0 \\
6049367398109200256 & 1 & 6.3 & 1.4 \\
6050172068822858624 & 1 & 0.0 & 0.7 \\
6050204641855013376 & 1 & 3.9 & 1.3 \\
6050345001390940160 & 2 & 0.6 & 0.8 \\
6050487456864139392 & 2 & 0.9 & 1.6 \\
6050626201483958400 & 1 & 2.0 & 2.3 \\
6050644102907664640 & 1 & 1.9 & 0.9 \\
6050681211424993280 & 1 & 4.3 & 2.1 \\
6051734990243252096 & 1 & 2.2 & 1.9 \\
6051764573978499200 & 2 & 0.8 & 0.9 \\
\hline
\end{tabular}
\caption{Objects in the common sample with published \Av and \gaia DR2 \Ag values (see Sect.~\ref{sect_cmd}). References are coded
as (1): \cite{Dunham_2015}, (2): \cite{2015MNRAS.448.2737R}.}
\label{tab:ap_tb_3}
\end{center}
\end{table}

\begin{landscape}
\begin{table}
\begin{tabular}{ccccccccccccc}
\hline
source\_id & G\_Flux & BP\_Flux & RP\_Flux & J\_Flux & H\_Flux & Ks\_Flux & W1\_Flux & W2\_Flux & W3\_Flux & W4\_Flux & SED\_Class & status \\
 & $\mathrm{W\,m^{-2}}$ & $\mathrm{W\,m^{-2}}$ & $\mathrm{W\,m^{-2}}$ & $\mathrm{W\,m^{-2}}$ & $\mathrm{W\,m^{-2}}$ & $\mathrm{W\,m^{-2}}$ & $\mathrm{W\,m^{-2}}$ & $\mathrm{W\,m^{-2}}$ & $\mathrm{W\,m^{-2}}$ & $\mathrm{W\,m^{-2}}$ &  &  \\
\hline
\hline
6046026394590627840 & 6.38E-15 & 1.62E-15 & 1.52E-14 & 3.75E-14 & 3.26E-14 & 2.17E-14 & 8.44E-15 & 5.24E-15 & 2.02E-15 & 1.68E-15 & Class  II & New \\
6047564611356188288 & 3.76E-15 & 4.11E-16 & 1.06E-14 & 7.34E-14 & 1.05E-13 & 8.79E-14 & 4.43E-14 & 2.88E-14 & 7.53E-15 & 5.38E-15 & Class  II & Control \\
6047569834036585728 & 4.24E-14 & 1.35E-14 & 9.17E-14 & 2.50E-13 & 2.80E-13 & 2.02E-13 & 9.91E-14 & 7.08E-14 & 3.23E-14 & 2.90E-14 & Class  II & Control \\
6047573849829430656 & 4.76E-16 & 1.25E-16 & 1.73E-15 & 2.22E-14 & 3.37E-14 & 3.57E-14 & 2.09E-14 & 1.64E-14 & 7.72E-15 & 5.90E-15 & Class  II & Control \\
6047584810585556480 & 2.84E-13 & 1.38E-13 & 5.05E-13 & 8.85E-13 & 9.02E-13 & 5.70E-13 & 2.10E-13 & 9.12E-14 & 6.38E-15 & 1.19E-15 & Photosphere & Other \\
6048982328524443136 & 1.37E-14 & 3.06E-15 & 3.17E-14 & 1.03E-13 & 1.21E-13 & 8.26E-14 & 3.21E-14 & 1.64E-14 & 5.10E-15 & 4.35E-15 & Class  II & New \\
6048993461079183488 & 1.90E-14 & 5.35E-15 & 4.19E-14 & 1.14E-13 & 1.27E-13 & 8.41E-14 & 3.05E-14 & 1.61E-14 & 1.06E-14 & 8.60E-15 & Class  II & New \\
6049011053265501056 & 8.30E-12 & 7.89E-12 & 9.48E-12 & 8.26E-12 & 5.20E-12 & 3.10E-12 & 1.04E-12 & 4.58E-13 & 2.62E-14 & 4.57E-15 & Photosphere & Other \\
6049043832452569728 & 5.25E-15 & 8.75E-16 & 1.33E-14 & 4.73E-14 & 4.34E-14 & 3.43E-14 & 1.55E-14 & 9.87E-15 & 2.83E-15 & 2.01E-15 & Class  II & New \\
6049045000683707392 & 5.56E-14 & 2.21E-14 & 1.09E-13 & 2.31E-13 & 2.31E-13 & 1.63E-13 & 7.03E-14 & 3.93E-14 & 1.03E-14 & 7.28E-15 & Class  II & Other \\
\hline
\end{tabular}
\caption{Extract from the photometry table. This entire table is available in the electronic version of this article.}
\label{tab:ap_tb_4}
\end{table}
\end{landscape}

\begin{figure*}[ht!]
  \centering \includegraphics[width=1.0\textwidth, trim = 0 50 0 0]{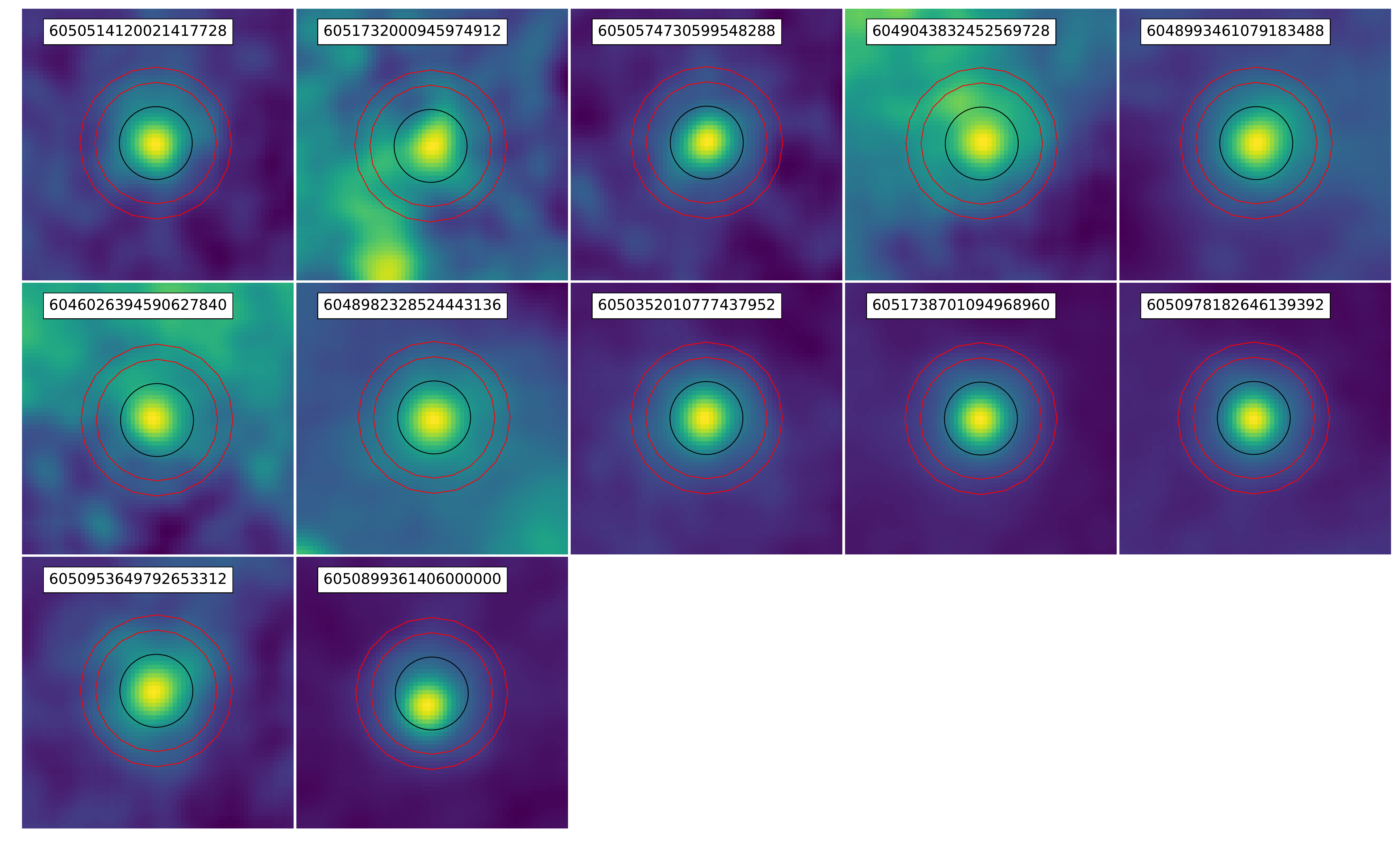}
  \caption{WISE W4 maps of the 12 new discs shown in Fig.~\ref{fig:fig_12}. The black line encompasses the aperture mask used to
  find the emission peak associated with the disc ($r = 12\arcsec$), and the red lines indicate the ring area used to estimate the background
  emission ($r_{in} = 20\arcsec$, $r_{out} = 25\arcsec$; see Sect.~\ref{sect_discs}). All the images are $1.5\arcmin \times 1.5\arcmin$ in
  size, and the white panels indicate the \gaia DR2 source ID.}
   \label{fig:ap_5}
\end{figure*}

\end{appendix}

%
%
\end{document}